\documentstyle[prl,preprint,aps,epsfig]{revtex}
\tightenlines
\begin{document}

\newcommand{\adag}{a^{\dag}}
\newcommand{\atil}{\tilde{a}}

\draft
\title{Study of the $^{7}Be(p,\gamma )^{8}B$ and $^{7}Li(n,\gamma )^{8}Li$
capture reactions using the shell model embedded in the continuum}

\author{K. Bennaceur\dag, F. Nowacki\dag\ddag,  J. Oko{\l}owicz\dag\S ~and
M. P{\l}oszajczak\dag}               
\address{\dag\ Grand Acc\'{e}l\'{e}rateur National d'Ions Lourds (GANIL),
CEA/DSM -- CNRS/IN2P3, BP 5027, F-14076 Caen Cedex 05, France}
\address{\ddag\ Laboratoire de Physique Th\'{e}orique  Strasbourg (EP 106),
3-5 rue de l'Universite, F-67084 Strasbourg Cedex, France }
\address{\S\ Institute of Nuclear Physics, Radzikowskiego 152,
PL - 31342 Krakow, Poland}


\maketitle

\begin{abstract}
\parbox{14cm}{\rm We
apply the realistic shell model which
includes the coupling between many-particle
(quasi-)bound states and the continuum of one-particle scattering states to
the spectroscopy of mirror nuclei: $^{8}$B and $^{8}$Li, as well as
to the description of low energy
cross sections (the astrophysical $S$ factors) in the capture reactions:
$^7\mbox{Be}(p,\gamma)^8\mbox{B}$ and
$^7\mbox{Li}(n,\gamma)^8\mbox{Li}$~. }

\end{abstract}
\bigskip

\vfill
\newpage

\section{Introduction}
The solution of solar neutrino problem, {\it i.e.}, an observed deficit of
neutrinos with respect to predictions of the Standard Solar Model (SSM)
\cite{bahcall} , is passing through an understanding of the capture reaction:
$^7\mbox{Be}(p,\gamma)^8\mbox{B}$. The $^{8}\mbox{B}$ produced in the solar
interior in
the reaction $^7\mbox{Be}(p,\gamma)^8\mbox{B}$, is the principal source of
high energy neutrinos detected in solar neutrino experiments. Therefore,
if we stay within the framework of standard electroweak theory and
the SSM \cite{bahcall1}, the observed deficit of those neutrinos can be
traced back, at least partially, to the value of low energy
$^7\mbox{Be}(p,\gamma)^8\mbox{B}$ radiative capture cross section
which determines the magnitude of neutrino flux and
remains the most uncertain quantity in the SSM \cite{bahcall1}.
At the solar energies
( $\sim 20\,$keV), this cross-section is too small to be directly measurable.
For this reason, the theoretical analysis of this reaction is so important.
 On the other hand, whenever the measurement is feasible
($E_{CM} > 150\,$keV), the
exact value of the capture cross section
depends : (i) on the normalization obtained indirectly from the
$^7\mbox{Li}(d,p)^8\mbox{Li}$ cross section and, (ii) on
the model dependent
extrapolation of measured values of the cross-section  down to the interesting
domain of solar energies. Experimental data
for the $^7\mbox{Be}(p,\gamma)^8\mbox{B}$ capture cross section are varying
strongly \cite{parker,kavanagh,vaughn,filippone,hammache} , though more
recent experiments consistently indicate low values ($S < 20$ eV$\cdot$b)
for the astrophysical factor $S \equiv
{\sigma}_{CM}(E_{CM}) E_{CM}\exp (-2\pi \eta)$, where $\eta =
e^{2}Z_1Z_2/{\hbar}v$~\cite{filippone,hammache}. Also the Coulomb
dissociation experiments \cite{motobayashi}
deduce low value for the astrophysical $S$ - factor
though this value depends strongly on the amount of $E2$ - contribution in
the Coulomb dissociation which is not yet completely understood.

Many different theoretical models have been applied for the
calculation of $S$ - factor at stellar energies
\cite{tombrello,aurdal,robertson,descouvement1,descouvement2,csoto,brown,nunes},
and their predictions are often in striking disagreement among each other,
confirming strong model and/or approximation dependence
of calculated cross-section. Part of the
theoretical ambiguities can be removed by a simultaneous study of the
$^7\mbox{Li}(n,\gamma)^8\mbox{Li}$ mirror reaction, which has also been
studied by several experimental groups \cite{imhof,wiescher,nagai}.
Also for this reaction, largely different
values for the low energy capture cross
section have been reported \cite{imhof,wiescher,nagai}. In the context of the
solar neutrino problem, the
$^7\mbox{Li}(n,\gamma)^8\mbox{Li}$ cross section is often used to
extrapolate the capture cross section for the reaction
$^7\mbox{Be}(p,\gamma)^8\mbox{B}$ down to the solar energies at
$E_{CM} \sim 20\,$keV \cite{filippone} .
But the $^7\mbox{Li}(n,\gamma)^8\mbox{Li}$ reaction cross section
at very low energies is also extremely interesting in itself as it
provides the essential element of rapid process of
primordial nucleosynthesis of nuclei with $A \geq 12$ in the inhomogeneous
big-bang models \cite{applegate,fuller,malaney,terasawa}.
Indeed, in the inhomogeneous
big-bang hypothesis \cite{applegate,fuller,malaney,terasawa},
the main reaction chain leading to the synthesis of heavy elements
is~\cite{malaney} $^1\mbox{H}(n,\gamma)^2\mbox{H}(n,\gamma)^3\mbox{H}(d,n)
^4\mbox{He}(t,\gamma)^7\mbox{Li}(n,\gamma)^8\mbox{Li}$,
and then
$^8\mbox{Li}(\alpha,n)^{11}\mbox{B}(n,\gamma)^{12}\mbox{B}
({\beta}^{-})^{12}C(n,\gamma)^{13}\mbox{C}$,
{\it etc.}, for heavier nuclei. In this sense,
the reaction $^7\mbox{Li}(n,\gamma)^8\mbox{Li}$  is a key process
to bridge the gap of mass $A=8$ and to produce heavy elements.

The theoretical description of weakly bound exotic
nuclei close to the drip-line, such as, {\it e.g.}, $^8\mbox{B}$ on the
proton rich side of
the drip line or $^{11}\mbox{Be}$, $^{11}\mbox{Li}$ on the neutron rich side of
the drip line,  is challenging due to the proximity of particle
continuum which implies the strong modification of effective nucleon - nucleon
interaction and causes the unusual spatial properties
(halo structures, a large diffusivity) of nucleon density
distribution . Nowadays, these properties are perhaps
somewhat better understood near the neutron drip line
than near the proton drip line. In weakly bound exotic nuclei,
number of excited bound states or narrow resonances
is small and , moreover, they couple strongly to the particle continuum.
Hence, these systems should be described in
the quantum open system formalism which does not artificially separate
the subspaces of (quasi-) bound (the $Q$-subspace) and scattering (the
$P$-subspace) states. For well bound nuclei
close to the $\beta$ - stability line,
microscopic description of states in the first subspace is given
by the nuclear shell model (SM) with model-space dependent
effective two-body interactions, whereas the latter subspace
is treated in terms of the coupled channels equations.
For weakly bound exotic nuclei, the validity of this
basic paradigm is certainly questionable, and we propose to
change it by considering the approximation which
takes into account  coupling between $Q$~ and $P$~ subspaces
in terms of the residual nucleon--nucleon interaction.
This coupling will consistently
modify both the scattering solutions and the spectroscopic quantities for
interior bound states.

One possibility for such an approach could be the Continuum Shell Model (CSM),
which in the
restricted space of configurations generated using the finite-depth potential
has been studied for the giant resonances and for the radiative capture
reactions probing the microscopic structure of these resonances
\cite{bartz1,bartz2,fladt,bartz3} . This approach may be
insufficient for the description of low lying excitations, in particular
in the nuclei close to drip lines, where it is essential to
have a most realistic description of bound states.
For that reason, the corner-stone
of our approach, which is called in the following
the Shell Model Embedded in the Continuum (SMEC), is
the {\it realistic} SM itself which is used
to generate the $N$-particle wave functions. This choice implies that
coupling between the SM
states and the one-particle scattering continuum must be given by the residual
nucleon - nucleon interaction. The first application of the SMEC model
has been published recently in Ref. \cite{bnop1} .

As said before, we are interested in
describing low lying bound and resonance states in exotic nuclei.
For that reason, we restrict the description of particle continuum to the
subset of one-nucleon decay channels. This should be a reasonable starting
point for describing both the microscopic structure of
$^8\mbox{B}$ and $^8\mbox{Li}$ and the corresponding reactions:
$^7\mbox{Be}(p,\gamma)^8\mbox{B}$ and $^7\mbox{Li}(n,\gamma)^8\mbox{Li}$.
At higher energies,
{\it e.g.}, above $\alpha$ or $t$ emission thresholds,
the one-particle continuum approximation
prohibits a reliable description of more complicated multi-nucleon
decay channels as well as the residual correlations generated
in the many-body wave functions of bound states
by the coupling to those channels.
In principle, this obstacle can be removed in future studies and
further improvement
of the SMEC (CSM) to include more complicated channels like, {\it e.g.},
the $\alpha$ - decay channel, can be done following the approach of
Balashov et al. \cite{balashov} . This rather involved extension of the SMEC
will not be discussed here any further. One should also be aware that
in the case of two-nucleon halo nuclei or at higher energies above the
threshold for the three-particle decay,
the one-particle continuum approximation is oversimplified.

The paper is constructed in the following way.
In Sect. II, we present some details of the SMEC model, stressing in particular
those elements which differ this model from the CSM.
Sect. III is devoted to
the discussion of spectroscopic properties of $^8\mbox{B}$ (Sect. III.B) and
$^8\mbox{Li}$ (Sect. III.C). We shall discuss influence of the residual
interaction which couples $Q$
and $P$ subspaces on the energy and the width of
many-body states. We will also discuss an important
self-consistency correction  to the average single particle (s.p.) potential
which results from this residual interaction. The self-consistent
average potential is then
used to generate radial formfactors of many-body states which enter the
coupling matrix elements between $Q$ and $P$ subspaces.
Sect. IV is devoted to the discussion of $^7\mbox{Be}(p,\gamma)^8\mbox{B}$ and
$^7\mbox{Li}(n,\gamma)^8\mbox{Li}$  capture cross sections. Finally,
the most important conclusions are summarized in Sect. V.

\section{The shell model embedded in the continuum (SMEC)}

In the decaying nucleus or in the reaction processes such as, {\it e.g.}, the
radiative capture
process, all asymptotic channels are given by the solutions with outgoing
waves. Such physical systems, in which bound (localized) interior states
are coupled to the asymptotic scattering channels,
are called the open systems. One way out of this complication is to
describe the quantum open system {\it together} with its
environment of asymptotic channels as
the quantum closed system. This implies that
the projection operator technique is used to separate $P$ subspace of
asymptotic channels from the $Q$ subspace of many-body states which are
build up by the bound s.p.\ wave functions
and by the s.p.\ resonance wave functions. In this latter case, one should use
certain cutoff radius to define this part of
resonance wave function which belongs to the $Q$ subspace and the remaining
part is put into the $P$ subspace.
In our case, the $P$ subspace
contains $(N-1)$ - particle states with nucleons on bound orbits
and one nucleon in the scattering state, but also this part of the s.p.\
resonance wave functions which is outside of the cutoff radius.
In fact, we define the $P$ subspace from the condition
\begin{eqnarray}
\label{pq}
P + Q = 1 ~ \ ,
\end{eqnarray}
which also implies that
all wave functions used in the calculations are
orthonormalized in the usual sense. Physically, the closeness
assumption for the total system means that
the states of $(N-1)$ - nucleus which define the
asymptotic channels for the studied reaction are assumed to be stable.
This assumption in turn means
that the most important channels are supposed to be
those which involve low-lying states of the
residual nucleus. In the domain of low energy excitations and/or for weakly
bound nuclei away from the $\beta $-stability line this is most likely a
good approximation. However, in the domain of higher
excitation like, {\it e.g.}, for the giant resonances, the quality of this
approximation which depends on the width of considered many-body states
in $N-1$ nucleus, may be hazardous for broad states.

The key element of the CSM is the treatment of
s.p.\ resonances, which on one side
may have an important amplitude inside a nucleus and, on other
side, they exhibit asymptotic behaviour of scattering wave functions
\cite{bartz1} .  The part of
resonance for $r < R_{cut}$~, where $R_{cut}$~ is the cut-off radius,
is included in $Q$ subspace,
whereas the remaining part is left in the $P$ subspace \cite{bartz1}. The
wave functions of both subspaces are then renormalized
in order to ensure the
orthogonality of wave functions in both subspaces. It should be mentioned that
in the earlier formulation of the CSM \cite{mahaux},
the contribution of the s.p.\ resonances both to $Q$ and $P$ subspaces
have been neglected.

In the SMEC calculations, we solve similar equations as in the CSM
\cite{bartz1} but , as explained below, due to specificity of exotic
nuclei, we change certain important ingredients of the CSM.
For the bound states we solve the standard SM problem:
\begin{eqnarray}
\label{eq1}
H_{QQ}{\Phi}_i = E_i{\Phi}_i ~ \ .
\end{eqnarray}
$H_{QQ} \equiv QHQ$~ is {\it identified} with
the SM Hamiltonian and its ${\Phi}_i$~ are the $N$-particle
(quasi-) bound wave functions. We believe, that for
a quantitative description of low
lying states in the exotic nuclei one has to use as a starting point the
accurate many-body wave
functions in the $Q$ subspace which are provided by the SM with effective
interactions.

The residual coupling of $P$ and $Q$ subspaces is given by
the zero-range force
\begin{eqnarray}
\label{force}
V_{12} = -V_{12}^{(0)}
[\alpha + (1-\alpha )P_{12}^{\sigma}]\delta({\bf r}_1 - {\bf r}_2) ~ \ ,
\end{eqnarray}
where $P_{12}^{\sigma}$ is the spin exchange operator.
We assume that the SM Hamiltonian $H_{QQ}$ contains already effects generated
by this residual coupling  and we do not modify $H_{QQ}$ anymore.

For the continuum part, we solve the coupled channel equations:
\begin{eqnarray}
\label{esp}
(E^{(+)} - H_{PP}){\xi}_{E}^{c(+)} = 0 ~  \ ,
\end{eqnarray}
where index $c$~ denotes
different channels and $H_{PP} \equiv PHP$~.
The superscript $(+)$ in (\ref{esp}) means that boundary
conditions  for outgoing scattering waves are used.
In our case, we have ingoing wave in the input channel
and outgoing waves in all other channels.  The structure of
$(N - 1)$~- nucleus is given by the SM, whereas one nucleon
occupies a scattering state. The channel states are defined by coupling one
nucleon in the continuum to a many-body state of $(N - 1)$ - nucleus.

The SM wave function has an
incorrect asymptotic radial behaviour for unbound states.
Therefore, to generate both the radial s.p.\ wave functions in the $Q$ space
and the scattering wave functions in $P$ space
we use the finite-depth average potential of Saxon-Woods (SW) type
with the spin-orbit part included:
\begin{eqnarray}
\label{pot}
U(r) = V_0f(r) + V_{SO} (4{\bf l}\cdot{\bf s})
\frac{1}{r}\frac{df(r)}{dr}  + V_C  ~ \ ,
\end{eqnarray}
where $f(r)$ is the spherical symmetric SW formfactor:
\begin{eqnarray}
\label{pot1}
f(r) = \left[ 1 + \exp \left( \frac{r-R_0}{a} \right) \right]^{-1} ~ \ .
\end{eqnarray}
The Coulomb potential $V_C$ is calculated for the uniformly charged sphere
with radius $R_0$.

Coupled channel eqs. (\ref{esp}) can be written more explicitly as:
\begin{eqnarray}
\label{esp1}
\sum_{c^{'}}^{}(E^{(+)} - H_{cc^{'}}) {\xi}_E^{c^{'}(+)} = 0 ~ \ ,
\end{eqnarray}
where
\begin{eqnarray}
\label{esp2}
H_{cc^{'}} = (T + U ){\delta}_{cc^{'}} + {\upsilon }_{cc^{'}}^{J} ~ \ .
\end{eqnarray}
In the above equation, $T$ stands for the kinetic-energy operator and
$ {\upsilon }_{cc^{'}}^{J}$ is the channel-channel coupling generated by the
residual interaction. The explicit formula
for $ {\upsilon }_{cc^{'}}^{J}$ is given in (\ref{form7}). The s.p.\ potential
in (\ref{esp2}) consists of an 'initial guess' $U(r)$, and the diagonal
part of the coupling potential ${\upsilon }_{cc}^{J}$
which depends on both the s.p.\ orbit
${\phi}_{l,j}$ and the considered many-body
state $J^{\pi}$. Obviously, this
correction cannot be neglected when generating s.p.\ wave function
${\phi}_{l,j}$ for a given $J^{\pi}$. These s.p.\ wave functions
define $Q$ subspace and thus modify the diagonal part of the residual force.
So this implies  a self-consistent iterative
procedure, because the change of s.p.\ wave
function changes in turn the correction coming from the residual force
(see the Appendix A). As long
as this correction is small in comparison with the initial average potential,
the iterative procedure is fastly converging
to the new self-consistent average potential:
\begin{eqnarray}
\label{usc}
U^{(sc)}(r) = U(r)+{\upsilon }_{cc}^{J(sc)}(r) ~ \ .
\end{eqnarray}
The parameters of the initial average potential (\ref{pot})
are chosen in such a way that the resulting potential $U^{(sc)}(r)$
reproduces energies of experimental s.p.\ states,
whenever their identification is
possible. In the next section, we will show examples of such self-consistent
potentials. The dependence of the generated correction in the potential
(\ref{usc}) on isospin of s.p.\ states
generates a mutual dependence of average potentials for protons and neutrons
in the iterative procedure. Using self-consistently determined radial
wave functions instead of those generated by $U(r)$ means that the matrix
elements $V_{\alpha \beta \gamma \delta}$  (A1)
of the residual force $V_{12}$ depend not only on the s.p.\ wave
functions ${\phi}_{l,j}$ involved but also on the many-body state
$J^{\pi}$. This is another interesting aspect of the
self-consistent procedure determining the average potential in SMEC.

The third system of equations in CSM consists of inhomogeneous
coupled channel equations:
\begin{eqnarray}
\label{coup}
(E^{(+)} - H_{PP}){\omega}_{i}^{(+)} = H_{PQ}{\Phi}_i \equiv w_i
\end{eqnarray}
with the source term $w_i$ which is primarily given by the SM
structure of $N$ - particle wave function for state ${\Phi}_i$~. The explicit
form of the source term is given in (\ref{form10}).
These equations define functions ${\omega}_{i}^{(+)}$~,  which
describe the decay of quasi-bound state ${\Phi}_i$~ in the continuum.
The source $w_i$~ couples the wave function of $N$ - nucleon
localized states with $(N-1)$ - nucleon localized states plus one nucleon
in the continuum. For zero-range residual force (\ref{force}),
formfactor of the source term is given by the s.p.\ wave functions of the
same self-consistently determined average
potential $U^{(sc)}(r)$ as used to define the subspaces $Q$ and~$P$.

The full solution of the many-body problem can be expressed by
means of three distinct functions:
${\Phi}_i$~, ${\xi}_{E}^{c}$~ and ${\omega}_i$ , and reads \cite{bartz1}~:
\begin{eqnarray}
\label{eq2}
{\Psi}_{E}^{c} = {\xi}_{E}^{c} + \sum_{i,j}({\Phi}_i + {\omega}_i)
\frac{1}{E - H_{QQ}^{eff}}
<{\Phi}_{j}\mid H_{QP} \mid{\xi}_{E}^{c}> ~ \ ,
\end{eqnarray}
where
\begin{eqnarray}
\label{eq2a}
H_{QQ}^{eff} = H_{QQ} + H_{QP}G_{P}^{(+)}H_{PQ} ~ \ ,
\end{eqnarray}
is the effective
SM Hamiltonian including the coupling to the continuum, and
$G_{P}^{(+)}$~ is the Green function for the motion of s.p.\ in
$P$ subspace. Matrix $H_{QQ}^{eff}$ is non-Hermitian
(symmetric and complex matrix)  for energies above the particle
emission  threshold
and Hermitian (real) for lower energies. Consequently,
the eigenvalues of $H_{QQ}^{eff}$ are real for bound states and
complex for decaying states. Matrix $H_{QQ}^{eff}$ can be diagonalized by
the orthogonal transformation:
\begin{eqnarray}
\label{transf}
{\Phi}_i \longrightarrow {\tilde {\Phi}_j} = {\sum}_{i}^{} b_{ji}{\Phi}_i ~ \
,
\end{eqnarray}
with complex eigenvalues:
${\tilde {E_i}} - \frac{1}{2}i{\tilde {{\Gamma}_i} }$ ,
which depend on the energy $E$ of particle in the continuum.
The eigenvalues of $H_{QQ}^{eff}$~ at energies  $\tilde{E_i}(E) = E$,
determine the
energies and widths of resonance states. With these changes, one obtains:
\begin{eqnarray}
\label{cons}
{\Psi}_{E}^{c} = {\xi}_{E}^{c} + \sum_{i}^{}{\tilde {\Omega}_i}
\frac{1}{E - {\tilde E_i}
+ (i/2){\tilde {\Gamma}_i}} <{\tilde {\Phi}_i} \mid H \mid {\xi}_{E}^{c}>
\end{eqnarray}
for the continuum many-body wave function projected on channel $c$ , where
\begin{eqnarray}
\label{diss}
{\tilde {\Omega}_i} = {\tilde {\Phi}_i} + \sum_{c}
\int_{{\varepsilon}_c}^{\infty} dE^{'} {\xi}_{E^{'}}^{c}
\frac{1}{E^{(+)} - E^{'}}
<{\xi}_{E^{'}}^{c}\mid H \mid {\tilde {\Phi}_i}> ~ \ ,
\end{eqnarray}
is  the wave function of discrete state modified by the coupling to the
continuum states. Hence, this formalism if fully {\it symmetric} in
treating the
continuum and bound state parts of the solution :
${\Psi}_{E}^{c}$ (in Eq. (\ref{cons}))
represents the continuum state modified by the discrete states, and
${\tilde {\Omega}_i}$ (in Eq. (\ref{diss})) represents the discrete
state modified by the coupling to the continuum. More informations about
those features of SMEC which are the same as in the CSM,
can be found in the paper by Bartz et al. \cite{bartz1}.

\section{Spectroscopy of $^{8}\mbox{B}$ and $^{8}\mbox{Li}$ nuclei}

\subsection{The self-consistent average potential spanning the $Q$ subspace}

The essential element of SMEC approach is the construction of $Q$ -
subspace. As explained in
the previous section, this is obtained by an iterative procedure which
for a given initial average s.p.\ potential (\ref{pot})
and for a given residual two-body interaction ({\ref{force})
yields the self-consistent s.p.\ potential depending on the
s.p.\ wave function ${\phi}_{l,j}$, the total spin $J$ of the $N$-nucleon
system as well as on the one-body matrix elements of $(N-1)$ - nucleon
daughter system.
The parameters of different initial SW potentials for
[$\mbox{p} \bigotimes {}^{7}\mbox{Be}$] and
[$\mbox{n} \bigotimes {}^{7}\mbox{Li}$] systems, which are
used in this work are summarized in Table \ref{parameters}.
All these potentials have the same parameters of radius $R_0=2.4\,$fm, surface
diffuseness $a=0.52\,$fm, and spin-orbit coupling $V_{SO}=-4\,$MeV.
Cohen - Kurath (CK) interaction \cite{cohen} is used as a SM interaction.
The potentials of Table \ref{parameters}
have been obtained for different parameters ($1-\alpha$)
of the spin-exchange component in the residual interaction (\ref{force})
and for different s.p.\ orbits ${\phi}_{l,j}$
which correspond to energies ${\varepsilon}_{l,j}$ (see the second column in
Table~\ref{parameters}) in the self-consistently determined
potential $U^{(sc)}(r)$.  In all cases, the strength of the
residual interaction (\ref{force}) is $V_{12}^{(0)}=650\,$MeV$\cdot$fm$^{3}$.
This value  yields a reasonable compromise between the description of
${}^{8}\mbox{B}$ and ${}^{8}\mbox{Li}$ spectra and in particular their decay
widths. The dependence of the energy spectrum
of ${}^{8}\mbox{B}$ on $V_{12}^{(0)}$ will be presented in the Sect. III.B.

In Fig. 1 , we show examples of calculated potentials in $^8\mbox{B}$
for the proton s.p.\ orbital $1p_{3/2}$ in two different total spin states:
$J^{\pi} = 2^+$ and $J^{\pi} = 1^+$. The calculations
have been performed using the initial potential $U(r)$ for
$(1-\alpha)=0.27$, which is chosen in such a way that the
self-consistent potential $U^{(sc)}(r)$ yields
$1p_{3/2}$ proton s.p.\ orbit bound at the
experimental binding energy  of ground state (g.s.) $J^{\pi}=2_{1}^{+}$.
The same $U(r)$ is then taken both for  $2^{+}$ and $1^{+}$ states.
The spectroscopic factor of proton $1p_{3/2}$ s.p.\ state
in the g.s., is close to 1 for CK interaction \cite{cohen}.
This allows to identify position of proton $1p_{3/2}$
s.p.\ orbit in $J^{\pi}=2^{+}$ state, {\it i.e.}, we
demand that $U^{(sc)}(r)$  provides $1p_{3/2}$ s.p.\ state
at  $-137\,$keV. This choice,
as we shall see in Sect. IV.A, is essential for a quantitative description of
$^{7}\mbox{Be}(p,\gamma )^{8}\mbox{B}$ radiative capture cross-section.

The self-consistent
average potential $U^{(sc)}(r)$ (the solid line in the l.h.s.\ plot)  exhibits
for small $r$ a clear maximum  which is absent in the initial potential
$U(r)$ . One should also notice, that self-consistent
potentials $U^{(sc)}(2^{+})$ and $U^{(sc)}(1^{+})$ are different
(compare solid curves on l.h.s.\  and
r.h.s.\ of Fig. 1) , in spite of the fact that the initial potential
$U(r)$ in these states
is the same. This clearly shows strong
state dependence of calculated average fields and two-body effective
interactions. For example, whereas the $1p_{3/2}$ s.p.\ orbit in the
g.s.\ ($J^{\pi}=2_{1}^{+}$) is
at $-137\,$keV , the self-consistent procedure
 yields this orbit at $-2.538\,$MeV in the first excited state
($J^{\pi}=1_{1}^{+}$) .

The dotted lines in Fig. 1
show the equivalent s.p.\ average potentials $U^{(eq)}(r)$ in $J^{\pi}=2^{+}$
and $1^{+}$ states for proton $1p_{3/2}$ orbit.
For the same $a$, $R_0$ and $V_{SO}$ parameters
as in the initial potential $U(r)$, parameter $V_0$ in $U^{(eq)}(r)$
is adjusted in order  to  reproduce the energy of
$1p_{3/2}$ s.p.\ orbit in $U^{(sc)}(r)$.
Clearly, $U^{(eq)}(r)$ and $U^{(sc)}(r)$ differ strongly in the
potential interior.  However, the potential
surface of $U^{(eq)}(r)$ and $U^{(sc)}(r)$ for both considered $J^{\pi}$
is very similar. This is particularly apparent
for $J^{\pi}=2^{+}$. The root mean squared (r.m.s.) radius of the
$1p_{3/2}$ orbit is ${<r^2>}^{1/2}=4.228\,$fm in $U^{(sc)}(2^{+})$ and
${<r^2>}^{1/2}=4.239\,$fm in $U^{(eq)}(2^{+})$.
For $J^{\pi}=1^{+}$, the effective surface diffuseness of
$U^{(sc)}(r)$ is even slightly decreasing.
Also the radial wave functions of the proton $1p_{3/2}$ orbit in
$U^{(sc)}(r)$ and $U^{(eq)}(r)$  are almost identical with only
a small shift towards the potential interior of the maximum of
radial wave function in $U^{(sc)}(r)$.

We have no clear indication concerning
the position of proton $1p_{1/2}$ s.p.\ orbit. Using
the same $U(r)$ as used to determine $U^{(sc)}(r)$
for $1p_{3/2}$ s.p.\ state, we get
the $1p_{1/2}$ proton s.p.\ orbit in $U^{(sc)}(r)$ at
${\varepsilon}_{p_{1/2}}=+0.731\,$MeV
in $J^{\pi}=2^{+}$ states and
at ${\varepsilon}_{p_{1/2}}=+0.311\,$MeV in $J^{\pi}=1^{+}$ states .
Consequently, the energy splitting of $p_{3/2}$ and $p_{1/2}$ orbitals
is state dependent. It is $\Delta \varepsilon =3.369\,$MeV,
in the initial potential $U(r)$.
In $U^{(sc)}(r)$,  it equals
$\Delta \varepsilon =0.868\,$MeV for $J^{\pi}=2^{+}$,
and  $\Delta \varepsilon =2.849\,$MeV for $J^{\pi}=1^{+}$ states.
In $J^{\pi}=3^{+}$ states, coupling to the continuum does not introduce any
correction to the
position of $1p_{1/2}$ s.p.\ state. Therefore, to obtain the energy
and the wave function of $1p_{1/2}$ state
we take the equivalent potential $U^{(eq)}(3^{+})$ to $U^{(sc)}(3^{+})$
for the $1p_{3/2}$.

In general, the surface region of average potential shows weak sensitivity
 to the self-consistent correction. Deviations from this general rule can be
seen for weakly-bound many-body
states having an important admixture of
$l=0$ and $l=1$ neutron s.p.\ states, {\it i.e.}, for halo configurations in
the neutron rich nuclei \cite{riisager}.
In this case, the self-consistent correction may change significantly
the surface of the average potential \cite{fut}.
In the above discussion we have considered effect of coupling between
$P$ and $Q$ subspaces on the average s.p.\ potential for the
proton orbits. To exclude the
effect of Coulomb barrier, in Fig. 2 we show the initial
potential $U(r)$, the
self-consistent potential $U^{(sc)}(r)$, and the equivalent potential
$U^{(eq)}(r)$ for neutron $1p_{3/2}$ s.p.\ orbit in
$^{8}\mbox{Li}$ where, for the sake of argument,
$1p_{3/2}$ neutron s.p.\ orbit in $U^{(sc)}(r)$ is at $-20\,$keV.
Parameters of the initial potential in this case are:
$V_0=-33.4\,$MeV, $R_0=2.4\,$fm,
$a=0.52\,$fm and $V_{SO}=-4\,$MeV. Parameters of the residual interaction
(\ref{force}) are the same as used in Fig. 1.
One may notice that the self-consistent
correction to the average potential is now much stronger than in
$^{8}\mbox{B}$ and, in particular, it changes strongly the surface of
$U^{(sc)}(r)$, as expected for weakly-bound halo nuclei \cite{riisager}.

Fig. 3 shows the dependence of $U^{(sc)}(r)$ on the
parameter $(1-\alpha )$ of the spin-exchange component of the two-body
residual interaction (\ref{force}).
The self-consistent potentials for different values of $(1-\alpha)$
have been determined for the
$1p_{3/2}$ s.p.\ proton orbital (${\varepsilon}_{p_{3/2}}=-137\,$keV) in
$J^{\pi}=2^+$. Parameters of corresponding initial potentials $U(r)$ are
shown in Table~\ref{parameters}. We see that $U^{(sc)}(r)$
strongly depends on the the spin-exchange term of the residual force
(\ref{force}). The largest correction to the initial potential $U(r)$ is
obtained for small values of $(1-\alpha)$, {\it i.e.}, when approaching the
Wigner force limit. This dependence of $U^{(sc)}(r)$ on the spin-exchange
component of the residual interaction
has a sensible effect on the width of resonances, as we shall see below.

\subsection{Spectrum of $^8\mbox{B}$}

In this section and in Sect. III.C, we shall present the detailed
calculations of spectra for mirror nuclei : $^8\mbox{B}$ and
$^8\mbox{Li}$. The SMEC results depend on the following ingredients :
(i) the nucleon - nucleon interaction in $Q$ subspace,
(ii) the residual coupling of
$Q$ and $P$ subspaces, (iii) the self-consistent
average s.p.\ potential which generates the radial
formfactor for s.p.\ bound wave functions and s.p.\ resonances,
and (iv) the cutoff radius for s.p.\ resonances.
The freedom of choosing
asymptotic conditions in solving eqs. (\ref{coup}) means
that a zero of excitation energy scale can be fixed
arbitrarily. In the following, we choose the
zero on this energy scale by requiring, independently
for any SMEC calculation shown in Fig. 4 and in Tables \ref{depv0} and
\ref{b8rcut10},  that the lowest experimentally known resonance
($J^{\pi}=1_{1}^{+}$ state for $^8$B)
with respect to the proton emission threshold has its energy equal to its
experimental value.
The same convention is used in the
calculation of the $^{7}\mbox{Be}(p,\gamma )^{8}\mbox{B}$
capture cross section in Sect. IV.A.

Fig. 4 compares SM energy spectrum for $T=1$~ states of $^{8}\mbox{B}$~
(on the l.h.s.) calculated in the $p$ - shell using the
CK interaction \cite{cohen}~,
with those obtained in the SMEC in different approximations.
The experimental data are plotted on the r.h.s.\ of this figure.
For the spin-exchange parameter,
we take $(1-\alpha) =0.27$, which is a standard value resulting from a fit
to the giant dipole resonance in $^{16}\mbox{O}$~\cite{bartz1,buck,rotter}~.
Different sets of coupling matrix elements are taken into account. In the
column denoted 'SMEC', we include coupling matrix elements
between the $J^{\pi}=3/2^{-}$ g.s.\ wave function of $^{7}\mbox{Be}$ and
all considered states in $^{8}\mbox{B}$. 'SMEC1' and 'SMEC1$^{*}$' columns
show the results when also coupling matrix elements
between the $J^{\pi}=1/2^{-}$ first excited  state in
$^{7}\mbox{Be}$ and states of
$^{8}\mbox{B}$ are included. In 'SMEC1', this $J^{\pi}=1/2^{-}$ state
is at the energy ($E^{*}=1.07\,$MeV) predicted by SM,  which differs
from the experimental energy ($E^{*}=0.429\,$MeV) of this level.
For that reason, in the column denoted 'SMEC1$^{*}$',
we show results of calculations where energy of
$J^{\pi}=1/2^{-}$ state is put at its
experimental position, without changing neither the
coupling matrix elements of residual force nor the effective interaction
in $Q$ subspace.

The iterative procedure to correct $U(r)$ and  to
include the diagonal part contribution of residual interaction has been
described in the previous section.
The self-consistently determined s.p.\ potential $U^{(sc)}(r)$ is then
used to calculate the radial formfactors of coupling matrix elements
and the s.p.\  wave functions.  One and the same
potential $U(r)$ is used for the calculation of self-consistent
potentials for all many-body states in $^{8}\mbox{B}$, and for both
$1p_{3/2}$ and $1p_{1/2}$ proton s.p.\ states.
These self-consistent
potentials for $1p_{3/2}$ proton s.p.\ state in $J^{\pi}=2^{+}$ and $1^{+}$
states of $^{8}\mbox{B}$ can be seen in Fig. 1.
For neutrons, there is no correction from the residual interaction,
and the average s.p.\ potential is chosen such that it
yields $1p_{3/2}$  and $1p_{1/2}$ neutron orbits
at $-13.02\,$MeV and  $-11.16\,$MeV respectively. These values have
been deduced from experimental $Q$-values and SM spectroscopic factors.
The parameters of initial potential for neutrons are :
$V_0=-67.595\,$MeV (for $1p_{3/2}$ state) and $V_0=-70.942\,$MeV
(for $1p_{1/2}$ state). The remaining parameters are the same as for protons.
The r.m.s.\ radius of neutron $1p_{3/2}$ and $1p_{1/2}$ orbitals is
${<r^2>}^{1/2} \simeq 2.5\,$fm.
We have checked that this choice of potential parameters for neutrons
is not crucial and, {\it e.g.}, shift of neutron
s.p.\ energies by ${}\sim 2\,$MeV does not influence final results of SMEC
for the spectroscopy of $^{8}\mbox{B}$.
Supplementary informations concerning
results shown in Fig. 4 can be found in Table~\ref{b8rcut10}.

The dependence of the $^{8}\mbox{B}$ energy spectrum on the strength
$V_{12}^{(0)}$ of the residual interaction (\ref{force}) is shown in Table
\ref{depv0}. The SMEC calculations correspond to the standard value
of the spin-exchange parameter:
$(1-\alpha )=0.27$. The calculated energy spectrum depends in a
complicated way on the strength parameter $V_{12}^{(0)}$. The optimal value of
$V_{12}^{(0)}$, in particular in what concerns the width of resonance
states which provide a most sensible test of the wave functions and the
values of matrix elements of residual coupling,
should be searched in the interval between $600$ and $700$ MeV$\cdot$fm$^3$.
In the following, in all SMEC calculations presented in this work we use
$V_{12}^{(0)} = 650$ MeV$\cdot$fm$^3$, and discuss in some details the
dependence of $^{8}\mbox{B}$ and $^{8}\mbox{Li}$ spectra on the spin-exchange
term in the residual interaction (\ref{force}).

The spectrum of $^{8}\mbox{B}$ is relatively insensitive to certain
approximations in SMEC. Ground state
energy relative to the proton emission threshold is reasonably well
reproduced by the SMEC.
The position of g.s.\ with respect to the
energy threshold for proton emission changes by about 100 keV due to the
inclusion of coupling to the g.s.\ of $^{7}\mbox{Be}$
(compare columns denoted 'SM' and 'SMEC' in
Fig. 4 and in Table~\ref{b8rcut10}). The coupling to the excited
$^{7}\mbox{Be}^{*}$ nucleus is relatively unimportant (see columns denoted
'SMEC1' and 'SMEC1$^{*}$' in Fig. 4 and in Table~\ref{b8rcut10}).
Calculated width of $J^{\pi}=1_{1}^{+}$ depends
weakly on chosen couplings  (compare 'SMEC' with 'SMEC1' or
'SMEC1$^*$' columns in Table~\ref{b8rcut10}) and for the interaction with
$(1-\alpha)=0.27$, one obtains approximately half of the experimental width.
Width for $J^{\pi}=1_{1}^{+}$ state depends
sensitively on the proportion of direct and
spin-exchange terms (see Table~\ref{b8adep}) in the residual
interaction, and the  agreement with the experimental value improves
when approaching the limit of Wigner force
($\alpha = 1$).
In this limit, the residual interaction $V_{12}$ in (\ref{force})
is compatible with SU(4) supermultiplet symmetry \cite{wigner}.
It is well known that this limit is badly broken in majority of heavier nuclei
mainly due to the increasing importance of spin-orbit coupling.
However, in $p$-shell nuclei, SU(4) is still
a good approximation \cite{nang,mukhopadhyay}. The Gamow-Teller transitions
have been studied using the SM calculations with CK interaction showing that
the supermultiplet symmetry of SU(4) is well preserved \cite{mukhopadhyay}.
Also realistic SM
calculations have confirmed that g.s.\ wave functions have very small
admixture of SU(4) - breaking configuration , both at the beginning and at the
end of $p$ - shell \cite{cohen,nang,john}.
For that reason, it is encouraging that the residual two-body
interaction between $Q$ and $P$, which is used in
SMEC calculations for low-lying many body resonances in $^{8}\mbox{B}$,
is consistent with the approximate validity of SU(4) scheme found
in the SM.

It is instructive to see the energy dependence of $J^{\pi}=1_{1}^{+}$ lowest
eigenvalue of the effective Hamiltonian (\ref{eq2a}) in $^{8}\mbox{B}$
(see Fig. 5) . The same interaction and potentials as in Fig. 4 are used.
Results shown with the solid line
include the coupling of $J^{\pi}=1_{1}^{+}$ state
to the channel wave function obtained by the coupling of
$J^{\pi}=3/2^{-}$ g.s.\ wave function of $^{7}\mbox{Be}$ with
one proton scattering wave function.
The dashed line shows results obtained by including couplings both
to the g.s.\ and to the $J^{\pi}=1/2^{-}$ first excited state of
$^{7}\mbox{Be}$ at the energy given by the SM calculation.
Finally, the dotted line shows the results when this first
excited state is placed at the
experimental position ($E^{*}=0.429\,$MeV). The real part of the eigenvalue
depends on the total energy of the system. For the center of mass (c.m.)
energies less than about 2 MeV,
$E_R$ is a decreasing function of energy and for higher energies it becomes an
increasing function of energy. The minimum of $E_R(E)$
depends slightly on whether the
coupling to the $J^{\pi}=1/2^{-}$ first excited state of $^{7}\mbox{Be}$ is
included. On the contrary, the imaginary part of the eigenvalue increases
 monotonously with energy from the threshold and saturates at higher energies.
Different approximations concerning the couplings have an
influence on $E_R$ and are practically insignificant for ${\Gamma}_R$.

SM energy of the first $J^{\pi}=3_{1}^{+}$ level is
too low as compared to the experimental value (see Fig. 4 and
Table \ref{b8adep}).
The coupling to the continuum cannot correct for this deficiency of SM.
The width of $3_{1}^{+}$ state differs by at least a factor 5
from the experimental data and here, again, the agreement between experiment
and calculations improves when
$\alpha \rightarrow 1$  (see Table~\ref{b8adep}).
There may be several reason for the observed
discrepancy between SMEC calculations and the experimental data for this
state.  Firstly, as mentioned already, SM is not well describing energy of
this state. (The CK interaction is somewhat less successful in describing
spectra of light $p$-shell nuclei.) Secondly, the wave function of
experimental $3_{1}^{+}$ state is most probably overlapping with the cluster
configuration [${^{3}\mbox{He}}-{^{5}\mbox{Li}}$], which cannot be adequately
described in $p$-space SM calculations. It is known
in mirror nucleus $^{8}\mbox{Li}$,
 that  the $3_{1}^{+}$ state is strongly influenced
by the mirror cluster configuration
[${^{3}\mbox{H}} - {^{5}\mbox{He}}$] \cite{stowe}~.  Moreover, energy of
experimental $3_{1}^{+}$ state lies above the threshold
for three-particle decay:
$^{8}\mbox{B} \longrightarrow [{^{3}\mbox{He}}-\mbox{p}-{^{4}\mbox{He}}]$.
This decay channel which surely contributes to the experimental width,
cannot be taken into account in the one-particle approximation for
scattering continuum. Also correlations which could be
generated in $Q$ subspace by the coupling to this three-particle
decay channel are outside of considered model space.
Thirdly, $3_{1}^{+}$ state in
$^{8}\mbox{B}$ cannot couple to the first excited state $1/2^{-}$ of
$^{7}\mbox{Be}$ but could couple to higher excited, particle-unstable
states such as $7/2^{-}$. These couplings  cannot be treated fully
consistently in SMEC, because the
closeness condition (\ref{pq}) implies that states in ($N - 1$) - nucleus
must be stable with respect to the particle emission.

\subsection{Spectrum of $^8\mbox{Li}$}

Table~\ref{li8} compares SM energy spectrum with those obtained in the SMEC
for $T=1$~ states of $^{8}\mbox{Li}$.
SM calculations in the $p$ - shell ($Q$ subspace) are done using
CK interaction \cite{cohen}, like in the case of $^{8}\mbox{B}$.
Parameters of the
residual interaction (\ref{force}) are also the same as given in caption of
Table~\ref{b8rcut10}. Parameters of $U(r)$  are given in
Table~\ref{parameters}.
$U(r)$ is such that the neutron  $1p_{3/2}$ s.p.\ state in $U^{(sc)}(r)$
is bound by $2.033\,$MeV in $J^{\pi}=2_{1}^{+}$ state, similarly as the
experimental g.s.\ of $^{8}\mbox{Li}$. The $1p_{1/2}$
neutron state is then bound by $1.109\,$MeV.  The r.m.s.
of $1p_{3/2}$ neutron s.p.\ orbit in $U^{(sc)}(r)$ is
${<r^2>}^{1/2} \simeq 3.474\,$fm.  The same
initial potential is used for calculation of self-consistent average
potentials for all other many-body states in $^{8}\mbox{Li}$ and for both
$1p_{3/2}$ and $1p_{1/2}$ neutron s.p.\ states.

Experimentally, $3_{1}^{+}$ state in $^{8}\mbox{Li}$
lies above the neutron emission threshold and we fix
zero on the energy scale by requiring that for all
different examples of SMEC calculations shown in Tables~\ref{li8} and
\ref{li8adep}, the energy position of this lowest resonance
with respect to the neutron emission threshold
corresponds to the experimental value. The same convention is used also in the
calculation of the $^{7}\mbox{Li}(n,\gamma )^{8}\mbox{Li}$
capture cross section in Sect. IV.B.

Coupling to the continuum
induces the renormalization of spin-orbit interaction.
The energy splitting of $1p_{3/2}$ and $1p_{1/2}$ orbitals
which equals : $\Delta \varepsilon =2.302\,$MeV
in $U(r)$, becomes $\Delta \varepsilon =0.924\,$MeV
in $U^{(sc)}(2^{+})$, and $\Delta \varepsilon =3.078\,$MeV
in $U^{(sc)}(1^{+})$ .  One should notice that these
splittings are {\it different}
from analogous splittings in $^{8}\mbox{B}$ (see the discussion in Sect 3.B).
Hence, the coupling to the continuum
breaks explicitly the mirror symmetry of
spectra and wave functions of SM.

The dependence of self-consistent correction to the average potential on
isospin of s.p.\ states, induces a salient dependence of neutron
s.p.\ potential on proton s.p.\ potential. This
dependence is very weak in the studied cases of $^{8}\mbox{Li}$ and
$^{8}\mbox{B}$. For protons, $U(r)$
is such that it yields $1p_{3/2}$ orbit at $-14.8\,$MeV
and $1p_{1/2}$ orbit at $-13.9\,$MeV. These values have
been deduced, similarly as for $^{8}\mbox{B}$, from the experimental
$Q$-values and SM spectroscopic factors. The parameters of $U(r)$ are
$V_0=-73.066\,$MeV (for $1p_{3/2}$ state) and $V_0=-78.058\,$MeV
(for $1p_{1/2}$ state), and remaining parameters are the same as
for neutrons.

In the column denoted 'SMEC' in Table~\ref{li8}, we include the coupling
between $J^{\pi}=3/2^{-}$ g.s.\ of $^{7}\mbox{Li}$ and
all considered states of $^{8}\mbox{Li}$. 'SMEC1' and 'SMEC1$^{*}$' columns
show the results when also the coupling
between $J^{\pi}=1/2^{-}$ first excited state in
$^{7}\mbox{Li}$ and states of
$^{8}\mbox{Li}$ are included. In 'SMEC1', this $J^{\pi}=1/2^{-}$ state
is at the SM energy : $E^{*}=1.07\,$MeV. This energy value differs
from the experimental value : $E^{*}=0.467\,$MeV.
For that reason,
in 'SMEC1$^{*}$' column we show results where the energy of
$J^{\pi}=1/2^{-}$ state is placed at its experimental position. The
coupling matrix elements of the residual force (\ref{force})
and the CK interaction in $Q$
subspace remain unchanged in 'SMEC1' and 'SMEC1$^{*}$' calculations.

The position of g.s.\ with respect to the
energy threshold for neutron emission is weakly modified by the inclusion of
coupling to the continuum. The energy shift in this case is smaller than the
shift caused by the coupling to $J^{\pi}=1/2^{-}$ excited state  of
$^{7}\mbox{Li}^{*}$. In general, the coupling of $^{8}\mbox{Li}$ to
$^{7}\mbox{Li}^{*}$ plays more important role than the analogous coupling
of $^{8}\mbox{B}$ to $^{7}\mbox{Be}^{*}$ .
 Ground
state energy with respect to the neutron emission threshold is not well
reproduced by SMEC calculations. Energies of lowest three states :
$J^{\pi}= 2_{1}^{+}$, $1_{1}^{+}$, $3_{1}^{+}$ in SM calculation, are
too much compressed  as  compared to the data.
On the other hand, the energy splitting between $3_{1}^{+}$ state  and
the second excited state $1_{2}^{+}$ is well reproduced.
The width of $3_{1}^{+}$ state does not depend
on the chosen coupling matrix elements
and is by factor $\sim 10$ too small as compared to the data.
The width of $J^{\pi}=1_{2}^{+}$ state
depends also weakly on the coupling matrix
elements (see columns denoted by 'SMEC1' and 'SMEC1$^{*}$') and is
comparable to the experimental decay width.

Table \ref{li8adep} compares SM energy spectrum for
$T=1$~ states of $^{8}\mbox{Li}$~
(on the l.h.s.) with those obtained in the SMEC
for different parameters $(1-\alpha )$
of the residual interaction (\ref{force}).
A satisfactory agreement with the data for the width of $1_{2}^{+}$ state
is obtained when approaching the limit of Wigner force ($\alpha = 1$).
Also the width of $3_{1}^{+}$ state
improves largely and differs only by factor 3 from the data. We consider that
this agreement is satisfactory in view of the obvious limitations :
(i) SM is not well describing energy of
this state, (ii) the wave function
of the $3_{1}^{+}$ state is strongly overlapping  with the cluster
configuration $[^{3}\mbox{H} - {^{5}\mbox{He}}]$ which cannot be reliably
described in $p$ - shell calculations \cite{stowe}, and
(iii) $3_{1}^{+}$ state cannot couple to the first excited state $1/2^{-}$ of
$^{7}\mbox{Li}$ but could couple to higher excited states, such as
$7/2^{-}$,  which are particle unstable and  cannot be
included in the SMEC model.  One should notice that SMEC describes better
$3_{1}^{+}$ in $^{8}\mbox{Li}$ than in $^{8}\mbox{B}$ due to the fact
that the three-particle decay channel in $^{8}\mbox{Li}$ is closed.

Contrary to the $J^{\pi}=3_{1}^{+}$ state,
the $J^{\pi}=1_{2}^{+}$ state in $^{8}\mbox{Li}$ is well
reproduced, in particular, when the parameter of
spin-exchange component in the residual interaction (\ref{force})
approaches the limit of the Wigner force.
It is interesting to see the energy dependence of second
eigenvalue $1_{2}^{+}$ of the effective Hamiltonian (\ref{eq2a}).
The results shown with the solid line in Fig. 6
include the coupling of $J^{\pi}=1_{2}^{+}$ in $^{8}\mbox{Li}$ to the
$J^{\pi}=3/2^{-}$ g.s.\ of $^{7}\mbox{Li}$. One neutron is in the continuum.
The dashed line presents results obtained by including
couplings both
to the g.s.\ and to the $J^{\pi}=1/2^{-}$ first excited state of
$^{7}\mbox{Li}$ at the energy predicted by the SM. Finally, the dotted
line shows the results when this first
excited state is placed at the experimental position $E^{*}=0.467\,$MeV.
Contrary to the the case of $J^{\pi}=1_{1}^{+}$ lowest eigenvalue in
$^{8}\mbox{B}$, now also the imaginary part ${\Gamma}_R$ of the eigenvalue
depends on the chosen couplings.

\subsection{The quadrupole moment of $^{8}\mbox{B}$ and $^{8}\mbox{Li}$}

The quadrupole moment $<Q>$ of $^{8}\mbox{B}$ provides a useful test of
the SMEC wave function, in particular of its radial part. It
can be calculated following the approach of Carchidi et al
\cite{carchidi}, which has been recently applied by Brown at al \cite{brown} to
the case of $^{8}\mbox{B}$ and $^{8}\mbox{Li}$. The quadrupole moment
($ \sim ~<r^2Y^{(2)}>$), is obtained by summing over
products of many-body matrix elements and s.p. matrix elements \cite{carchidi}:
\begin{eqnarray}
\label{qmom}
<J_i||r^2Y^{(2)}||J_i> = \sum_{j,j^{'},t_z}^{}\frac{1}
{\sqrt 5}<J_i||[a_{j,t_z}^{+} {\tilde a}_{j^{'},t_z}]^{(2)}||J_i>
<j,t_z||r^2Y^{(2)}||j^{'},t_z> ~~~ \ ,
\end{eqnarray}
where the sum over $t_z$ runs over protons and neutrons. Inserting the complete
set of states of the $A=7$ system between the operators $a_{j,t_z}^{+}$
and ${\tilde a}_{j^{'},t_z}$, one can write (\ref{qmom}) as a sum over all
$p$ - shell states of $A=7$ system \cite{millener1} with $t_z = p, n$ , for
$^{7}\mbox{Be}$ and $^{7}\mbox{B}$ respectively. Consequently, one can express
the matrix element (\ref{qmom}) in terms of corresponding
spectroscopic amplitudes, effective charges,
geometrical factors and radial matrix elements \cite{brown}.
Different radial matrix
elements which enter the final expression for $<Q>$ (see eq. (20) of Ref.
\cite{brown} ) correspond to the terms coming from
the g.s. of $^{7}\mbox{Be}$ (the so-called
valence term : $<r^2>_v^{(p)}$), and the excited states in $^{7}\mbox{Be}$
(the $p$ - core proton term : $<r^2>_{p-c}^{(p)}$) and in $^{7}\mbox{B}$
(the $p$ - core neutron term : $<r^2>_{p-c}^{(n)}$), respectively
\cite{brown,carchidi}. For our
radial formfactors, calculated using the self-consistently determined average
potential for the spin-exchange term with $(1-\alpha )=0.05$,
the radial matrix elements are
: $<r^2>_v^{(p)} = 17.9~ \mbox{fm}^2$,
$<r^2>_{p-c}^{(p)} = 7.65 ~\mbox{fm}^2$ and
$<r^2>_{p-c}^{(n)} = 6.03 ~\mbox{fm}^2$. The position of $p$ - core proton and
neutron states have been deduced from the experimental $Q$ - values.
With the above values of radial matrix elements ,
one finds : $<Q>=6.99 ~\mbox{e fm}^2$, in the good agreement with the
experimental value \cite{minamisono} : $<Q> = 6.83 \pm 0.21 ~\mbox{e fm}^2$.
This theoretical value, which is dominated by the valence term,
has been obtained assuming the effective charges : $e_p=1.35$ ,
$e_n=0.35$, and the SM spectroscopic factors for the CK interaction.
The analogous calculation in $^{8}\mbox{Li}$ yields :
$<Q> = 2.78 ~\mbox{e fm}^2$, close to the experimental value
\cite{minamisono} : $<Q> = 3.27 \pm 0.06 ~\mbox{e fm}^2$. In this case,
the calculated radial matrix elements are :
$<r^2>_v^{(n)} = 12.1 ~\mbox{fm}^2$,
$<r^2>_{p-c}^{(n)} = 7.45 ~\mbox{fm}^2$ and
$<r^2>_{p-c}^{(p)} = 5.9 ~\mbox{fm}^2$, respectively. Even though the
agreement between those
theoretical estimates of $<Q>$ in $^{8}\mbox{B}$ and $^{8}\mbox{Li}$ and
the corresponding experimental values is encouraging, nevertheless it
should not be overstated in view of some uncertainties
concerning values of the effective charges,
the spectroscopic factors
and the radial matrix elements for the core terms
$<r^2>_{p-c}^{(p)}$ and $<r^2>_{p-c}^{(n)}$.

\section{Radiative capture processes involving $^{8}\mbox{B}$ and
$^{8}\mbox{Li}$ nuclei in the final state}

\subsection{The $^{7}\mbox{Be}(p,\gamma)^{8}\mbox{B}$~ reaction}

The ${\beta}^{+}$~ decay of $^{8}\mbox{B}$~ , which is formed by the reaction
$^{7}\mbox{Be}(p,\gamma)^{8}\mbox{B}$~
at the c.m.\ energy of about 20 keV,
is the main source of high energy solar neutrinos. In the absence of agreement
between different experimental data for this reaction and in view of the
disagreement between different measurements of solar neutrinos,
the input of SSM \cite{bahcall}  should be compared with the theoretical
values. The cross-section for the reaction
$^{7}\mbox{Be}(p,\gamma)^{8}\mbox{B}$~ remains the main uncertainty in the
input of the SSM.

In the SMEC, the initial wave
${\Psi}_i$~ for the system $[^{7}\mbox{Be} + \mbox{p}]_{J_i^{\pi}}$ is :
\begin{eqnarray}
\label{psiin}
\Psi_i(r)=\sum_{l_a j_a}i^{l_a}{\psi_{l_a j_a}^{J_i}(r)\over r}
\biggl[\bigl[Y^{l_a}\times\chi^{s}\bigr]^{j_a}\times\chi^{I_t}\biggr]^{(J_i)}
_{m_i}
\end{eqnarray}
and the final wave ${\Phi}_f$ for the system
$[^{8}\mbox{B}]_{J_f^{\pi}=2^{+}}$~ is:
\begin{eqnarray}
\label{psifin}
\Psi_f(r)=\sum_{l_b j_b}A_{l_bsj_b}^{j_bI_bJ_f} {u_{l_b j_b}^{J_f}(r)\over r}
\biggl[\bigl[Y^{l_b}\times\chi^{s}\bigr]^{j_b}\times\chi^{I_t}\biggr]^{(J_f)}
_{m_f}  \ .
\end{eqnarray}
$I_t$~ and $s$~ denote the spin of target nucleus and
incoming proton, respectively.
$A_{l_bsj_b}^{j_bI_bJ_f}$~ is the coefficient of fractional parentage and
$u_{l_bj_b}^{J_f}$~ is the s.p.\ wave in the many-particle state~$J_f$~.
With the wave ${\Psi}_i(r)$~ and ${\Psi}_f(r)$~,
we calculate the transition amplitudes:
\begin{eqnarray}
\label{x1}
\matrix{
\qquad T^{E{\cal L}} = C(E{\cal L})i^{l_a} \hat J_f \hat l_b \hat j_b\hat j_a
  <{\cal L} \delta J_f m_f \mid J_i m_i>
  <l_b 0 {\cal L} 0 \mid l_a 0> \hfill\cr
  \hfill\times W(j_b I_t {\cal L} J_i J_f j_a)
  W(l_b s {\cal L} j_a j_b l_a)
  I_{l_a j_a, l_b j_b}^{{\cal L},J_i} \qquad\cr}
\end{eqnarray}
for $E1$~ and $E2$~ and :
\begin{eqnarray}
\label{x2}
\matrix{
 T^{M1} = i^{l_a} \mu_N\hat J_f
  <1 \delta J_f m_f \mid J_i m_i> \hfill\cr
\qquad\times
\Biggl\{ W(j_b I_t 1 J_i;J_f j_a) \hat j_a \hat j_b \hfill\cr
  \hfill \Biggl[
    \mu\biggl({Z_t\over m_t^2}+{Z_a\over m_a^2}\biggr)
    \hat l_a \tilde l_a
    W(l_b s 1 j_a;j_b l_a)
+ (-1)^{jb-ja} 2 \mu_a \hat s  \tilde s
    W(s l_b 1 j_a;j_b s)\Biggr] \cr
\hfill + \mu_t (-1)^{J_f-J_i} \hat I_t \tilde I_t
 W(I_t j_b 1 J_i J_f I_t) \delta_{j_a j_b}
\Biggr\} \delta_{l_a l_b} I_{l_a j_a, l_b j_b}^{0,J_i} \cr}
\end{eqnarray}
for $M1$ transitions, respectively. In the above formula,
$\delta=m_i-m_f,\ \hat a \equiv \sqrt{2a+1},
\ \tilde a \equiv \sqrt{a(a+1)}$ and :
\begin{eqnarray}
\label{overlap}
I_{l_a j_a, l_b j_b}^{{\cal L},J_i} =\int u_{l_b j_b}
r^{\cal L}\psi_{l_a j_a}^{J_i} dr ~ \ .
\end{eqnarray}
The radiative capture cross section can then be expressed as:
\begin{eqnarray}
\label{tran}
\sigma^{E1,M1} = {16\pi\over9} \biggl({k_\gamma\over k_p}\biggr)^3
  \biggl({\mu\over\hbar c}\biggr)
  \biggl({e^{2}\over\hbar c}\biggr)
  {1\over 2s+1}~{1\over 2I_t+1}
\sum\mid T^{E1,M1}\mid^2
\end{eqnarray}
\begin{eqnarray}
\label{tran1}
\sigma^{E2} = {4\pi\over75} \biggl({k_\gamma^5\over k_p^3}\biggr)
  \biggl({\mu\over\hbar c}\biggr)
  \biggl({e^{2}\over\hbar c}\biggr)
  {1\over 2s+1}~{1\over 2I_t+1}
\sum\mid T^{E2}\mid^2
\end{eqnarray}
where $I_t$~ and $s$~ denote the spin of target nucleus and the spin of
incoming proton, respectively. $\mu$~ stands for the reduced mass of the
system.

Fig. 7 shows the calculated multipole contributions to the total
capture cross section as a function of
c.m.\ energy for different parameters of the residual interaction
(\ref{force}).  In the upper part of the figure,
the calculation is done for the same parameters of the residual
interaction as used in the calculations of spectra shown in Fig. 4. In the
lower part, the calculation is done for $(1-\alpha) = 0.05$,
close to the Wigner force limit. The spectrum in this case
can be seen in Table \ref{b8adep}.
Parameters of the initial potential $U(r)$ in these
two cases can be read from Table~\ref{parameters}.
The zero of the excitation energy scale
is chosen as described in Sect. III.B. With this convention,
the photon energy is given by the difference of
c.m.\ energy of $[\mbox{p} \bigotimes {^{7}\mbox{Be}}]_{J_{i}^{+}}$~ system
and the experimental energy of the $2_{1}^{+}$
g.s.\ of $^{8}\mbox{B}$. As can be realized from Fig. 7,
the $E1$ and $E2$ contributions as well as the total cross-section
are insensitive to the amount of spin-exchange
in the residual force. On the contrary, the $M1$ contribution and particularly
its resonant part, are strongly dependent on $\alpha$. Ratio of
$E2$ and $E1$ contributions at the position of $1_{1}^{+}$ resonance is
$8.15\cdot$$10^{-4}$ or $7.72\cdot$$10^{-4}$
depending on whether the spin-exchange parameter
equals 0.27 (the upper part of Fig. 7) or 0.05 (the lower part of Fig. 7).
The experimentally deduced value for this ratio
$6.7_{-1.9}^{+2.8}\cdot 10^{-4}$ \cite{davids} is consistent with our finding
but does not allow to distinguish between different spin-exchange parameters.

$E1$ component provides the main
 contribution to the total capture cross-section in the reaction
$^{7}\mbox{Be}(p,\gamma)^{8}\mbox{B}$~.
This non-resonant contribution is a good measure of the spatial
extension of $2_{1}^{+}$ wave function, which in turn is determined by the
extension of the proton $1p_{3/2}$ orbit. It is essential for the calculated
cross-section that the
$1p_{3/2}$ proton orbit in the self-consistent average potential is bound by
$137\,$keV. Modification of
this value by different choice of the depth parameter $V_0$ in $U(r)$,
introduces the change in $S^{E1}$ which can be much larger larger
than the change due to uncertainties
in the potential radius $R_0$ or its surface diffuseness $a$.

Fig. 8 shows the total $S$ - factor as a function of the c.m.\ energy.
The SMEC results correspond to $\alpha = 0.95$.
Different multipole contributions to the total cross section for this
parameter of the residual interaction (\ref{force}) have been
shown in the lower part of Fig. 7. Together with SMEC results for the
$S$-factor, we show experimental data \cite{filippone,hammache}.
The low energy dependence of $S(E)$ can be
fitted by \cite{descouvement2}:
\begin{eqnarray}
\label{fitproc}
S(E) = S(0) \exp ({\hat \alpha}E+{\hat \beta}E^{2}) ~ \ .
\end{eqnarray}
In our case, the fit of $S(E)$ using (\ref{fitproc})
in the range of c.m.\ energies up to
100 keV yields $S(0)=19.594$ eV$\cdot$b, ${\hat \alpha}=-1.544\,$MeV$^{-1}$,
${\hat \beta}=6.468\,$MeV$^{-2}$.
As compared to the similar fit of $S$ - factor calculated
in the Generator Coordinate Method \cite{descouvement2},
we find smaller $S(0)$ parameter and slightly
bigger values for the parameters ${\hat \alpha}$ and ${\hat \beta}$.
Recent experimental determination of
$S$-factor \cite{hammache} yields similar to ours low value for $S(0)$.

 The ratio of $M1$ and $E1$ contributions for $\alpha=0.95$ is:
${\sigma}^{M1}/{\sigma}^{E1}=1.43{\cdot}10^{-3}$, $2.65{\cdot}10^{-3}$ and
$1.90{\cdot}10^{-2}$~ at 20, 100 and 500 keV, respectively. The
resonant part of $M1$~ transitions
yields $S^{M1}=20.52$ eV${\cdot}$b at the $1_{1}^{+}$
resonance energy. This $M1$-contribution to the astrophysical $S$-factor
decreases fast and becomes $S^{M1}=3.65{\cdot}10^{-1}$,
$4.74{\cdot}10^{-2}$, $2.72{\cdot}10^{-2}$ eV${\cdot}$b
at c.m.\ energies 500, 100, 20 keV, respectively. At the
position of $1_{1}^{+}$ resonance, the calculated
$S$ - factor ($S = 40.67$ eV${\cdot}$b)
is smaller than measured by Filippone et al. \cite{filippone} .
This value , which is dominated by the $M1$-contribution, is proportional to
the square of spectroscopic amplitude of $p$-states,
which for the CK interaction is
$-0.352$ and 0.567 for $p_{1/2}$ and $p_{3/2}$ respectively. Similar small
values of spectroscopic amplitudes
are obtained for Kumar \cite{kumar} and PTBME \cite{julies}
interactions (see also \cite{brown} ).

The $E2$ contribution to the astrophysical factor was recently measured by
Kikuchi et al. \cite{motobayashi1} who finds:
$S^{E2} \simeq 0.0_{-0}^{+0.8}\,$meV$\cdot$b and
$S^{E2} \simeq 0.0_{-0}^{+3 (+3.6)}\,$meV$\cdot$b
in the energy intervals from 1.25 to 1.5 MeV
and from 1.5 to 1.75 MeV respectively. SMEC gives for this quantity
$52 - 53\,$meV$\cdot$b and $53 - 71\,$meV$\cdot$b, in these two energy
intervals respectively. Similar values for $S^{E1}$ have
been been found by Typel and Baur \cite{typel}.
These values are by factor 10 larger than those determined by
Kikuchi et al. \cite{motobayashi1} what remains a puzzle.

\subsection{The $^{7}\mbox{Li}(n,\gamma)^{8}\mbox{Li}$~ reaction}

This mirror reaction to the above considered capture reaction:
$^{7}\mbox{Be}(p,\gamma)^{8}\mbox{B}$~, together with a simultaneous
description of energy spectra and particle decay widths of $^{8}\mbox{B}$ and
$^{8}\mbox{Li}$, provides another stringent test for SMEC calculations. The
SM interaction and SM many-body wave functions ({\it e.g.} the spectroscopic
amplitudes)
are identical in both cases. The self-consistent one-body potentials
which take into account residual coupling of $Q$ and $P$ subspaces and which
determine the radial formfactors of s.p.\ wave functions used in the
calculation of matrix elements of the residual interaction (\ref{force}),
are optimized in the same way in $^{8}\mbox{B}$ and  in
$^{8}\mbox{Li}$. Finally, the parameters of direct and spin-exchange terms
in the residual interaction (\ref{force}) are also the same,
so the modification of coupling matrix elements in $^{8}\mbox{B}$ and
$^{8}\mbox{Li}$ is solely due to the different radial shape of s.p.\ wave
functions in the corresponding self-consistent potentials for different
$J^{\pi}$ of many-body states.
In the case of neutrons, the integral in Eq.~(\ref{overlap}) is sensitive
to the nuclear interior even in the low energy limit.  From elastic
scattering of
neutrons the scattering lengths $a_S$, where $S$ is the channel spin
($\vec S = \vec s + \vec I_t$, {\it c.f.} Eqs.~(\ref{psiin}) and
(\ref{psifin})), are known to be \cite{lynn}:
$a_1 = 0.87 \pm 0.07\,$fm and $a_2 = -3.63 \pm 0.05\,$fm.
So for the $s$-wave in the initial channel we adopted procedure of Barker
\cite{barker} of readjustment of appropriate
$s$-wave scattering potentials in order to reproduce these experimental values
of scattering lengths.

In Fig. 9 we show
different multipole contributions to the total
capture cross section as a function of
c.m.\ energy for different parameters $\alpha$
of the residual interaction (\ref{force}).
 The same parameters have been used for the mirror reaction:
$^{7}\mbox{Be}(p,\gamma)^{8}\mbox{B}$ (see Fig. 7). In the
lower part of Fig. 9, the calculation is done for the spin-exchange parameter
equal $(1-\alpha)=0.05$. The corresponding parameters of $U(r)$ can be found
in Table~\ref{parameters}. The choice of zero on the excitation energy
scale is the same as described in Sect. III.C. With this
convention, the photon energy in Fig. 9 is given by the difference of
c.m.\ energy of $[\mbox{n} \bigotimes {^{7}\mbox{Li}}]_{J_{i}^{+}}$~ system
and the experimental energy of the $2_{1}^{+}$~
g.s.\ of $^{8}\mbox{Li}$~. As can be seen in Fig. 9, the total cross-section
and the $E1$ contribution in particular,
are insensitive to the amount of spin-exchange
in the residual force. The $E2$ contribution shows a weak sensitivity to the
parameter $\alpha$ in the region of $1_{2}^{+}$ resonance.
The $M1$ contribution  and in particular
its resonant part, are strongly dependent on $\alpha$. At the thermal neutron
energies, $M1$ contributions for $(1-\alpha) = 0.23$ and 0.05 differ by
approximately one order of magnitude.

Like for the
mirror reaction $^{7}\mbox{Be}(p,\gamma)^{8}\mbox{B}$,
the dominant contribution to the total capture cross-section in
$^{7}\mbox{Li}(n,\gamma)^{8}\mbox{Li}$~ reaction is the $E1$ component.
Nevertheless, the $M1$ contribution in
$^{7}\mbox{Li}(n,\gamma)^{8}\mbox{Li}$~ is relatively more important, in
particular near the $3_{1}^{+}$ resonance.
This is partially due to the smaller extension of
$1p_{3/2}$ neutron s.p.\ orbit in the g.s.\ wave function of
$^{8}\mbox{Li}$ ($1p_{3/2}$ neutron s.p.\ orbit is bound by 2.033 MeV in the
g.s.\ of $^{8}\mbox{Li}$) as compared to the extension of
$1p_{3/2}$ proton s.p.\ orbit in the g.s.\ of $^{8}\mbox{B}$. This strong
binding
of $1p_{3/2}$ neutron state in $2^{+}$ states of $^{8}\mbox{Li}$ has also a
direct consequence on calculated radiative capture cross section which
becomes reduced, mainly its $E1$ component.
One should also
underline that in low-energy reaction such as
$^{7}\mbox{Li}(n,\gamma)^{8}\mbox{Li}$ ,
neutron  penetrates interior region of the potential and, therefore,
is more sensitive to the ratio of direct and spin-exchange terms in
(\ref{force}) and, indirectly, to the
modifications of the interior of average potential by
the coupling to the continuum (see Figs. 1 -- 3).

Fig. 10 shows the total neutron capture cross-section
as a function of the c.m.\ energy.
The calculation is done for the same parameters of residual interaction
(\ref{force}) as used in Fig. 8 for the mirror reaction
$^{7}\mbox{Be}(p,\gamma)^{8}\mbox{B}$~.
Together with the SMEC results, we show the data of Nagai et al.
\cite{nagai} which measured the cross-section for the
$\gamma$ - decay to the g.s.\ of
$^{8}\mbox{Li}$. The calculation fits well the data at this energy.
The low energy dependence of calculated
total neutron capture cross section can be
fitted by \cite{book}:
\begin{eqnarray}
\label{fit}
\sigma (E) = \left( \frac{{\mu}_n}{2E} \right)^{1/2} \left( s_0 + s_1E^{1/2}
+ s_2E + \cdots \right) ~ \ ,
\end{eqnarray}
where ${\mu}_n$ is the reduced mass of neutron
and $E$ is the c.m.\ energy in MeV.
In the energy interval up to 100 keV SMEC results
can be well fitted by three-parameter fit:
$s_0 = 11.517$, $s_1 = -2.145$, $s_2 = -11.636$, when ${\mu}_n$ is expresed
in a.m.u.

 The ratio of $M1$ and $E1$ contribution is
${\sigma}^{M1}/{\sigma}^{E1}=1.78{\cdot}10^{-3}$, $2.09{\cdot}10^{-2}$ and
$1.09$~ at 20, 100 and 200 keV, respectively.  The
resonant part of $M1$~ transitions which is overestimated in the calculation
due to small calculated width for this state,
yields the contribution of ${\sigma}^{M1}=305.1$ ${\mu}$b at the
$3_{1}^{+}$ resonance energy.
This contribution decreases fast and becomes ${\sigma}^{M1}= 13.8$,
$0.426$, $9.18{\cdot}10^{-2}$ ${\mu}$b
at c.m.\ energies 200, 100, 20 keV, respectively.

\section{Conclusions}

In this work we have applied SMEC model
for the microscopic description of
spectra in $^{8}\mbox{B}$ and $^{8}\mbox{Li}$, and
low-energy radiative capture cross sections in mirror reactions:
$^{7}\mbox{Be}(p,\gamma)^{8}\mbox{B}$ and
$^{7}\mbox{Li}(p,\gamma)^{8}\mbox{Li}$. The SMEC model, in which
realistic SM solutions for (quasi-)bound states are coupled to the
one-particle scattering continuum,  is a development of
CSM model~\cite{bartz1,bartz2} for the description of low excitation energy
properties of weakly bound nuclei. For that reason, we use a
realistic SM effective interaction in the $Q$ subspace and introduce
a residual zero-range force with the
spin-exchange included which couples $Q$ and $P$ subspaces. This
deliberate choice of interactions implies that
the finite-depth potential generating potentials in $P$ and matrix elements of
residual coupling ($Q$ subspace), has to be determined self-consistently.
The self-consistent iterative procedure yields
new state-dependent average potentials and consistent with them
new renormalized matrix elements of the
coupling force. These renormalized couplings and average potentials
are then consistently used in the calculations of
spectra and capture cross-sections, {\it i.e.}, both in $Q$ and $P$
subspaces. What should be taken for coupling between bound and scattering
states is in principle not known and we have decided to use a schematic
combination of Wigner and Bartlett forces. Varying the
parameter of the spin-exchange component for a fixed
intensity of the coupling, we came to the conclusion that most satisfactory
description of experimental data is achieved for small contribution of the
spin-exchange part, {\it i.e.}, approaching the limit of pure Wigner force.
This finding is consistent with the results of
SM which strongly suggest an approximate validity of $SU(4)$
symmetry in $p$-shell nuclei\cite{cohen,nang,mukhopadhyay,john}. This proves
also an intrinsic consistency in our model between the effective SM
interaction, in our case the CK
interaction, and the residual coupling between $Q$ and $P$ subspaces.

Simultaneous study of mirror system and reactions allows for a better
understanding of the role that play different approximations and parameters
in the model. The dependence of final results on radius,
diffusivity and spin-orbit coupling
parameters of the initial potential $U(r)$ is not terribly
important and they can be taken from any reasonable systematics. Coupling to
the excited configurations in the $(N - 1)$ - daughter nucleus is somewhat
more important in $^{8}\mbox{Li}$ than in $^{8}\mbox{B}$. However, this
coupling depends only on the wave function of daughter nucleus in the
excited state and is totally
insensitive to the exact energy position of these excited configurations.
On the contrary, the depth of $U(r)$ has
to be carefully adjusted so that the energy of s.p.\ orbit(s) in $U^{(sc)}(r)$
involved in the systems:
$[{n} \bigotimes (N-1)]$ and $[{p} \bigotimes (N-1)]$,
reflects the binding of g. s. in the nucleus $N$. This is very important for
any quantitative analysis of the reaction cross-section.
In the studied cases of
$^{8}\mbox{B}$ and $^{8}\mbox{Li}$ , the correct identification of this
s.p.\ orbit and hence the determination of an appropriate depth parameter in
$U(r)$ is simple because the SM spectroscopic factor in $2_{1}^{+}$
g.s.\ is close to 1. Different binding of
$^{8}\mbox{B}$ and $^{8}\mbox{Li}$ leads in SMEC model to different
$U^{(sc)}(r)$ for corresponding many-body states in mirror nuclei. This in
turn causes: (i) the breaking of
initial mirror symmetry in SM spectra of these nuclei
 and (ii) the different effective spin-orbit splitting in
self-consistent potentials for corresponding states in mirror nuclei.

SMEC model in the present form allows to describe microscopically the
coupling to one-nucleon continuum. More complicated decay
channels like, {\it e.g.}, those involving the $\alpha$ particle in the
continuum or more than two particles in the asymptotic states are beyond
the scope of this model. It is encouraging, however, that these possible
shortcomings in the description of decay channels, as we have shown on the
example of $3_{1}^{+}$ resonances, are so unambiguously reflected
in the calculated decay width for these states. In general, the decay
width is particularly sensitive to the details of the SM wave functions
involved and to the values of matrix elements of residual coupling so
they provide a sensible test of the quality of SMEC wave functions
and/or approximations involved.

The overall agreement between experimental data and SMEC
calculations
for studied nuclei is good
proving the internal consistency of model assumptions and parameters. The
astrophysical factor $S(0)$ for the reaction
$^{7}\mbox{Be}(p,\gamma)^{8}\mbox{B}$ equals 19.424 eV$\cdot$b which is
close to the values reported by Filippone et al. \cite{filippone} and
Hammache et al. \cite{hammache} but differs from many
earlier experiments as well as from many theoretical analysis
\cite{descouvement1,csoto,brown} . The low-energy dependence of $S(E)$ is
slightly different from the one found by Descouvement and Baye
\cite{descouvement1} which is sometimes used in extrapolating the experimental
results to the astrophysically relevant region. The calculated
ratio of $S^{E1}$ and $S^{E2}$ contributions in the region of $1_{1}^{+}$
resonance also agrees well with the data of Davids et al. \cite{davids}.
Surprisingly, at higher energies ($E > 1.25\,$MeV) the calculated $E2$
contribution differs by factor 10 from the reported value of Kikuchi et al.
\cite{motobayashi1}. The results of Refs. \cite{davids} and
\cite{motobayashi1} seem to us incompatible with each other.

The present studies have shown that SMEC results depend sensitively on very
small number of parameters. Some of them, like the parametrization of the
residual interaction which couples states in $Q$ and $P$ subspaces, has been
established in the present work. The others, related to the energy of
s.p.\ states which determine the radial wave function of many-body states, are
bound by the SM spectroscopic factors and experimental binding energy in
studied nuclei. This gives us a confidence that SMEC can have large predictive
power when applied to other $p$ - shell nuclei and to other capture
cross-sections in this region.

\vskip 1truecm

{\bf Acknowledgements}\\
We wish to express our gratitude to S. Dro\.zd\.z and I. Rotter for many
stimulating discussions during the course of development of the SMEC model.
We thank also P. Van Isacker for interesting
discussions. This work was partly supported by
KBN Grant No. 2 P03B 097 16 and the Grant No. 76044
of the French - Polish Cooperation.

\appendix

\section{The coupling potential and the source term}

To solve the coupled channel equations (\ref{esp}) in $P$ subspace, one has to
calculate the matrix elements $V_{\alpha \beta \gamma \delta}$ of the
residual interaction (\ref{force}) between states in $Q$ and $P$ subspaces.
For the zero-range force including the spin exchanges, which was
used in this work, we have:
\begin{eqnarray}
\label{form5}
{\cal V}^L_{\alpha \beta \gamma \delta}(r_1,r_2)=
{\cal Z}^L_{ \alpha \beta \gamma \delta}
\frac{1}{r^2}
\delta(r_1-r_2) ~ \ ,
\end{eqnarray}
where
\begin{eqnarray}
\label{form5ipol}
{\cal Z}^L_{\alpha \beta \gamma \delta}
&=&\frac{V_{12}^{(0)}}{4\pi}\left[(a-b)
{\cal M}^{L1}_{\alpha \beta \gamma \delta}
+(1-\delta_{\tau_\alpha\tau_\beta}\delta_{\tau_\gamma\tau_\delta})(a+b)
{\cal M}^{L0}_{\alpha \beta \gamma \delta}
\right]
\end{eqnarray}
and angular two-body matrix element with isospin $T$ can be expressed
as~\cite{brussaard}~:
\begin{eqnarray}
\label{form6}
{\cal M}^{LT}_{\alpha \beta \gamma \delta}
&=& -\frac{1}{4} \left(1+(-1)^{l_{\alpha}+l_{\beta}-l_{\gamma}-l_{\delta}}
 \right) \frac{{\hat j}_{\alpha}
{\hat j}_{\beta}{\hat j}_{\gamma}{\hat j}_{\delta}}{{\hat L}^2}
  \nonumber \\
& & \times
\left\{ (-1)^{j_{\beta}+j_{\delta}+l_{\beta}+l_{\delta}}
{< j_{\alpha} {-\frac{1}{2}}\: j_{\beta} {\frac{1}{2}} | L0 >}
{< j_{\gamma} {-\frac{1}{2}}\: j_{\delta} {\frac{1}{2}} | L0 >}
\left[1-(-1)^{L+T+l_{\gamma}+l_{\delta}}\right] \right . \nonumber \\
& & \left . -
{< j_{\alpha} {\frac{1}{2}}\: j_{\beta} {\frac{1}{2}} | L1 >}
{< j_{\gamma} {\frac{1}{2}}\: j_{\delta} {\frac{1}{2}} | L1 >}
\left[1+(-1)^T\right] \right\}~ \ .
\end{eqnarray}
Any symbol containing the hat, like {\it e.g.} ${\hat j}$, means:
${\hat j} \equiv {\sqrt {2j+1}}$.

The channel-channel coupling in (\ref{esp}), (\ref{coup})
is taken into account through the matrix elements of the type:

\begin{eqnarray}
\label{form7}
{{\upsilon }_{cc'}}^J = -\sum_{L}
       \hat{L} \cdot {\cal V}^{L}_{\alpha \beta \gamma \delta}
\langle (J_t j_{\alpha})_{J} ||
\left[ \left( \adag_{\alpha}\adag_{\beta}\right) ^{L}
          \cdot \left( \atil_{\gamma}\atil_{\delta}\right)^{L}  \right] ^{0}
          ||(J'_t j_{\gamma})_{J} \rangle
\end{eqnarray}
The  $N$-body matrix element in the above expression,  is rewritten as :
\begin{eqnarray}
 \sum_{K}
       \hat{L}^2 \cdot \hat{K}^2 \cdot
{\scriptsize \left\{  \begin{array}{ccc}
   j_{\alpha} &j_{\beta}  & L \\
   j_{\gamma} &j_{\delta} & L \\
       K      &    K      & 0 \\
         \end{array} \right\}} \cdot
\langle (J_t j_{\alpha})_{J} ||
\left[ \left( \adag_{\alpha}\atil_{\gamma}\right)^{K}
          \cdot \left( \adag_{\beta}\atil_{\delta}\right)^{K}  \right] ^{0}
          ||(J'_t j_{\gamma})_{J} \rangle ~ \ ,
\end{eqnarray}
and the reduced matrix element in this latter formula is then expressed as :
\begin{eqnarray}
        \hat{J}^2 \cdot
{\scriptsize \left\{  \begin{array}{ccc}
   J_t  & j_{\alpha}  & J \\
   J'_t & j_{\gamma}  & J \\
       K      &    K      & 0 \\
         \end{array} \right\}} \cdot
\langle J_t  ||
\left( \adag_{\beta}\atil_{\delta}\right)^{K}
          ||J'_t  \rangle \cdot
\langle j_{\alpha}  ||
 \left( \adag_{\alpha}\atil_{\gamma}\right)^{K}
          ||j_{\gamma}  \rangle                  ~ \ .
\end{eqnarray}
The diagonal parts of this operator induce corrections which renormalize
the s.p.\ average potential $U(r)$ (see Eq. (\ref{usc}) ).

The source term in the inhomogeneous eqs. (\ref{coup})
takes into account the couplings of the type :

\begin{eqnarray}
\label{form10}
w^{(i)} = -\sum_{L}
       \hat{L} \cdot V^{L}_{\alpha \beta \gamma \delta}
\langle (J_t j_{\alpha})_{J} ||
\left[ \left( \adag_{\alpha}\adag_{\beta}\right) ^{L}
          \cdot \left( \atil_{\gamma}\atil_{\delta}\right)^{L}  \right] ^{0}
          || \Phi^i_J \rangle \\
  = -\sum_{L}
       \hat{L} \cdot V^{L}_{\alpha \beta \gamma \delta}
\langle (J_t j_{\alpha})_{J} ||
\left[  \adag_{\alpha} \left( \adag_{\beta}
          \left( \atil_{\gamma}\atil_{\delta}\right)^{L}
       \right)^{j_\alpha} \right] ^{0}
          || \Phi^i_J \rangle
\end{eqnarray}
Again, the reduced matrix element in this expression is given
as a coupled product of the two contributions:

\begin{eqnarray}
        \hat{J}^2 \cdot
{\scriptsize \left\{  \begin{array}{ccc}
   J_t        & j_{\alpha}  & J \\
   J          & 0           & J \\
   j_{\alpha} & j_{\alpha}  & 0 \\
         \end{array} \right\}} \cdot
\langle j_{\alpha}  ||
 \adag_{\alpha}
 ||0  \rangle.
\langle J_t  ||
\left( \adag_{\beta}\left(\atil_{\gamma}\atil_{\delta}\right)^{L}
 \right)^{j_\alpha} || \Phi^i_J \rangle
\end{eqnarray}
This operator modifies both real and imaginary parts of eigenvalues of
$H_{QQ}^{eff}$ in Eq. (\ref{eq2a}), but does not change the s.p.\ average
potential $U(r)$.















\vfill

\newpage


\begin{table}[h]

\caption{Parameters of the initial potentials $U(r)$ (\protect\ref{pot})
used in the calculations of self-consistent potentials $U^{(sc)}(r)$ for
various parameters $(1-\alpha)$ of the spin-exchange term in the residual
interaction (\protect\ref{force}). $U^{(sc)}(r)$ are constructed for
various $T=1$, positive parity states in $^{8}\mbox{B}$ and
$^{8}\mbox{Li}$ and for
different single particle states: $1p_{3/2}$ and $1p_{1/2}$. For all cases
the radius of the potential is $R_0=2.4\,$fm, the diffuseness parameter is
$a=0.52\,$fm, and the spin-orbit orbit parameter is
$V_{SO}=-4\,$MeV. The strength of the
residual interaction (\protect\ref{force}) is
$V_{12}^{(0)} = 650\,$MeV$\cdot$fm$^3$, for all considered cases.}
\label{parameters}
\begin{center}
\begin{tabular}{|c|l|l|l|}
\hline
System & ${\varepsilon}_{p_{3/2}}$ [MeV] & $1-\alpha$ & $V_0$ [MeV] \\
\hline
[p $\bigotimes$ $^{7}$Be]  & $ -0.137 $ & $ 0.45 $ & $-42.140$ \\
                &            & $ 0.27 $ & $-40.045$ \\
                &            & $ 0.05 $ & $-37.660$ \\
\hline
[n $\bigotimes$ $^{7}$Li]  & $ -2.033 $ & $ 0.45 $ & $-41.683$ \\
                &            & $ 0.27 $ & $-39.555$ \\
                &            & $ 0.05 $ & $-36.905$ \\
\hline
\end{tabular}
\end{center}
\end{table}

\begin{table}[h]

\caption{Dependence of the spectrum of $^{8}\mbox{B}$ on the
strength $V_{12}^{(0)}$ of the residual interaction (\protect\ref{force}) and
for a fixed value of the parameter of the spin-exchange term :
$(1-\alpha) = 0.27$. For more details, see the description in the text.}
\label{depv0}
\begin{center}
\begin{tabular}{|c|c|c|c|c|c|c|}
\hline
Nucleus & $J^{\pi}$ & Energy [keV] & Width [keV] & Energy (exp.) [keV] &
Width (exp.) [keV] &$V_{12}^{(0)}$
[MeV$\cdot$fm$^3$] \\
\hline
$^{8}$B & $2^{+}$    & $ -443 $ & ---  & $ -137.5 \pm 1.0 $ & --- & 500 \\
        &            & $ -441 $ & ---  &                    &     & 600  \\
        &            & $ -356 $ & ---  &                    &     & 650  \\
        &            & $ -339 $ & ---  &                    &     & 700  \\
        &            & $ -166 $ & ---  &                    &     & 800  \\
        & $1^{+}$    & $ ~637 $ & 2.9   & ~$ 637 \pm 6 $   & $ ~37 \pm 5 $ & 500 \\
        &            & $ ~637 $ & 3.5   &               &              & 600 \\
        &            & $ ~637 $ & 16.5   &               &              & 650 \\
        &            & $ ~637 $ & 19   &               &              & 700 \\
        &            & $ ~637 $ & 360  &                  &              & 800 \\
        & $3^{+}$    & $ ~1229 $ & 8.8   & ~$2183 \pm 30 $ & $~350 \pm 40 $ & 500\\
        &            & $ ~1221 $ & 9.2 &                &              & 600 \\
        &            & $ ~1294 $ & 13.1 &                &              & 650 \\
        &            & $ ~1311 $ & 12.1 &                &              & 700 \\
        &            & $ ~1470 $ & 16.2 &               &              & 800 \\
        & $1^{+}$    & $ ~2279 $ & 32   & ~$not~known $   & ~$not~known $ & 500 \\
        &            & $ ~2253 $ & 42   &               &              & 600 \\
        &            & $ ~2153 $ & 240.2   &            &              & 650 \\
        &            & $ ~2128 $ & 262   &               &              & 700 \\
        &            & $ ~2300 $ & 763   &             &              & 800  \\
\hline
\end{tabular}
\end{center}
\end{table}

\begin{table}[h]

\caption{SM  energies and SMEC
energies and widths vs.\ experimental ones of $^8$B
nucleus. All units are in keV.
The proton separation energy is adjusted in order to reproduce
the energy of the lowest resonance state.  Different labels denote as follows:
'SMEC' -- only the ground state of $^7$Be was taken into account in all
couplings, 'SMEC1' -- coupling to the first excited state in $^7$Be was
included with $E^* = 1.07\,$MeV (SM value), 'SMEC1$^*$' -- the same with
excited $^7$Be state at $E^* = 0.429\,$MeV (experimental value). Parameters of
the residual interaction (\protect\ref{force}) are:
$V_{12}^{(0)} = 650\,$MeV$\cdot$fm$^3$, $\alpha = 0.73$.
The cut-off radius is $R_{cut}=5\,$fm except for the $p_{1/2}$ s.p.\ wave
function in $1_{1}^+$ many body states, which is in the continuum at about
$300\,$keV above the threshold and for which
larger cut-off was used $R_{cut} = 10\,$fm.
The numbers in parentheses are the
widths of $3_{1}^+$ state if this state would be
placed at the experimental energy. For
more details, see the Table~\protect\ref{parameters} and the description
in the text.}
\label{b8rcut10}
\begin{center}
\begin{tabular}{| c | r | r r | r r | r r | r @{$\pm$} l r @{$\pm$} l |}
\hline
State & \multicolumn{1}{ c |}{SM} & \multicolumn{2}{ c |}{SMEC} &
\multicolumn{2}{ c |}{SMEC1} & \multicolumn{2}{ c |}{SMEC1$^*$} &
\multicolumn{4}{ c |}{experiment}\\
J$^{\pi}$ & energy & energy & width & energy & width &  energy & width &
 \multicolumn{2}{ c }{energy} & \multicolumn{2}{ c |}{width} \\
\hline
$ 2^+ $ & $ -446 $ & $ -356 $ & --- & $ -334 $ & --- &  $ -329 $ & --- &
 $ -137.5 $ & $ 1.0 $ & \multicolumn{2}{ c |}{ --- } \\
$ 1^+ $ & $  637 $ & $  637 $ & $ 16.5 $ & $ 637 $ & $ 15.3 $ &
 $ 637 $ & $ 15.2 $ & $ 637 $ & $  6 $ & $  37 $ & $ 5 $ \\
$ 3^+ $ & $ 1246 $ & $ 1294 $ & $ 13.1 $ & $ 1237 $ & $ 12.5 $ &
 $ 1241 $ & $ 12.6 $ & $ 2183 $ & $ 30 $ & $ 350 $ & $ 40 $ \\
 & & & $ (25.2) $ & & $ (25.8) $ & & $ (25.8) $ & \multicolumn{4}{ c |}{} \\
$ 1^+ $ & $ 2327 $ & $ 2153 $ & $ 240.2 $ & $ 2081 $ & $ 272.7 $ &
 $ 2080 $ & $ 309.0 $ & \multicolumn{4}{ c |}{{\it not known}} \\
\hline
\end{tabular}
\end{center}
\end{table}

\begin{table}[h]

\caption{The dependence of $^8$B spectra on the relative strengths of
direct and spin exchange parts of the residual interaction
(\protect\ref{force}).
Analogously to the entry 'SMEC' in Table~\protect\ref{b8rcut10}
only ground state of $^7$Be was taken into account.
The entries in this table are labelled by
the value of $\alpha$~parameter of the residual force.
The numbers in parentheses as in Table~\protect\ref{b8rcut10}. For more
details see
Table~\protect\ref{parameters}, the caption of Table~\protect\ref{b8rcut10}
and the discussion in the text.}
\label{b8adep}
\begin{center}
\begin{tabular}{| c | r | r r | r r | r r | r @{$\pm$} l r @{$\pm$} l |}
\hline
State & \multicolumn{1}{ c |}{SM} &
\multicolumn{2}{ c |}{$\alpha = 0.55$} &
\multicolumn{2}{ c |}{$\alpha = 0.73$} &
\multicolumn{2}{ c |}{$\alpha = 0.95$} &
\multicolumn{4}{ c |}{experiment}\\
J$^{\pi}$ & energy & energy & width & energy & width &  energy & width &
 \multicolumn{2}{ c }{energy} & \multicolumn{2}{ c |}{width} \\
\hline
$ 2^+ $ & $ -446 $ & $ -418 $ & --- & $ -356 $ & --- &  $ -320 $ & --- &
 $ -137.5 $ & $ 1.0 $ & \multicolumn{2}{ c |}{ --- } \\
$ 1^+ $ & $  637 $ & $  637 $ & $ 12.6 $ & $ 637 $ & $ 16.5 $ &
 $ 637 $ & $ 25.9 $ & $ 637 $ & $  6 $ & $  37 $ & $ 5 $ \\
$ 3^+ $ & $ 1246 $ & $ 1313 $ & $ 3.0 $ & $ 1294 $ & $ 13.1 $ &
 $ 1275 $ & $ 34.9 $ & $ 2183 $ & $ 30 $ & $ 350 $ & $ 40 $ \\
 & & & $ (5.6) $ & & $ (25.2) $ & & $ (67.4) $ & \multicolumn{4}{ c |}{} \\
$ 1^+ $ & $ 2327 $ & $ 2299 $ & $ 115.9 $ & $ 2153 $ & $ 240.2 $ &
 $ 1899 $ & $ 398.9 $ & \multicolumn{4}{ c |}{{\it not known}} \\
\hline
\end{tabular}
\end{center}
\end{table}

\begin{table}[h]

\caption{The same as in Table~\protect\ref{b8rcut10} but for $^8$Li
nucleus.  The neutron separation energy is adjusted in order to reproduce
the energy of the lowest resonance state.  Different labels denote as follows:
'SMEC' -- only the ground state of $^7$Li was taken into account in all
couplings, 'SMEC1' -- coupling to the first excited state in $^7$Li was
included with $E^* = 1.07\,$MeV (SM value), 'SMEC1$^*$' -- the same with
excited $^7$Li state at $E^* = 0.467\,$MeV (experimental value).
The numbers in parentheses are the widths of $1_{2}^+$ state if this state
would be
placed at the  experimental energy. For other informations see
Table~\protect\ref{parameters}, the the caption of
Table~\protect\ref{b8rcut10} and the discussion in the text. }
\label{li8}
\begin{center}
\begin{tabular}{| c | r | r r | r r | r r | r @{$\pm$} l r @{$\pm$} l |}
\hline
State & \multicolumn{1}{ c |}{SM} & \multicolumn{2}{ c |}{SMEC} &
\multicolumn{2}{ c |}{SMEC1} & \multicolumn{2}{ c |}{SMEC1$^*$} &
\multicolumn{4}{ c |}{experiment}\\
J$^{\pi}$ & energy & energy & width & energy & width &  energy & width &
 \multicolumn{2}{ c }{energy} & \multicolumn{2}{ c |}{width} \\
\hline
$ 2^+ $ & $ -1471.1 $ & $ -1437.09 $ & --- & $ -1346.4 $ & --- &  $ -1345.6 $
 & --- & $ -2033.8 $ & $ 0.3 $ & \multicolumn{2}{ c |}{ --- } \\
$ 1^+ $ & $  -388.1 $ & $  -418.9 $ & --- & $ -349.5 $ & --- & $ -351.5 $
 & --- & $ -1053.0 $ & $  0.1 $ & \multicolumn{2}{ c |}{ --- } \\
$ 3^+ $ & $ 221.2 $ & $ 221.2 $ & $ 3.4 $ & $ 221.2 $ & $ 3.5 $ &
 $ 221.2 $ & $ 3.5 $ & $ 221.2 $ & $ 3.0 $ & $ 33 $ & $ 6 $ \\
$ 1^+ $ & $ 1301.6 $ & $ 1065.6 $ & $ 357.9 $ & $ 1050.0 $ & $ 383.1 $ &
 $ 1036.7 $ & $ 464.2 $ & \multicolumn{2}{ c }{$ 1176 $} &
 \multicolumn{2}{ c |}{$ \approx 1000$}\\
 & & \multicolumn{2}{ r |}{$(378.1)$} & \multicolumn{2}{ r |}{$(422.9)$} &
 \multicolumn{2}{ r |}{$(506.8)$} & \multicolumn{4}{ c |}{} \\
\hline
\end{tabular}
\end{center}
\end{table}

\begin{table}[h]
\caption{The dependence of $^8$Li spectroscopy on the relative strengths of
direct and spin exchange parts of the residual interaction is presented.
Analogously to the entry 'SMEC' in Table~\protect\ref{li8}, only g.s.\ of
the $^7$Li was taken into account. The entries in this table are labelled by
the value of $\alpha$~parameter of the residual force.
The numbers in parentheses as in Table~\protect\ref{li8}.
For other informations see Table~\protect\ref{parameters}, the the caption of
Table~\protect\ref{li8} and the discussion in the text. }
\label{li8adep}
\begin{center}
\begin{tabular}{| c | r | r r | r r | r r | r @{$\pm$} l r @{$\pm$} l |}
\hline
State & \multicolumn{1}{ c |}{SM} &
 \multicolumn{2}{ c |}{$\alpha = 0.55$} &
 \multicolumn{2}{ c |}{$\alpha = 0.73$} &
 \multicolumn{2}{ c |}{$\alpha = 0.95$} &
 \multicolumn{4}{ c |}{experiment}\\
J$^{\pi}$ & energy & energy & width & energy & width &  energy & width &
 \multicolumn{2}{ c }{energy} & \multicolumn{2}{ c |}{width} \\
\hline
$ 2^+ $ & $ -1471.1 $ & $ -1514.7 $ & --- & $ -1437.9 $ & --- &  $ -1330.3 $
 & --- & $ -2033.8 $ & $ 0.3 $ & \multicolumn{2}{ c |}{ --- } \\
$ 1^+ $ & $  -388.1 $ & $  -441.7 $ & --- & $ -418.9 $ & --- & $ -394.4 $
 & --- & $ -1053.0 $ & $  0.1 $ & \multicolumn{2}{ c |}{ --- } \\
$ 3^+ $ & $ 221.2 $ & $ 221.2 $ & $ 0.8 $ & $ 221.2 $ & $ 3.4 $ &
 $ 221.2 $ & $ 9.4 $ & $ 221.2 $ & $ 3.0 $ & $ 33 $ & $ 6 $ \\
$ 1^+ $ & $ 1301.6 $ & $ 1199.2 $ & $ 171.8 $ & $ 1065.6 $ & $ 357.9 $ &
 $ 790.0 $ & $ 561.1 $ & \multicolumn{2}{ c }{$ 1176 $} &
 \multicolumn{2}{ c |}{$ \approx 1000$}\\
 & & \multicolumn{2}{ r |}{$(170.1)$} & \multicolumn{2}{ r |}{$(378.1)$} &
 \multicolumn{2}{ r |}{$(735.3)$} & \multicolumn{4}{ c |}{} \\
\hline
\end{tabular}
\end{center}
\end{table}


\vfill
\newpage

\begin{figure}[t]
\centerline{\epsfig{figure=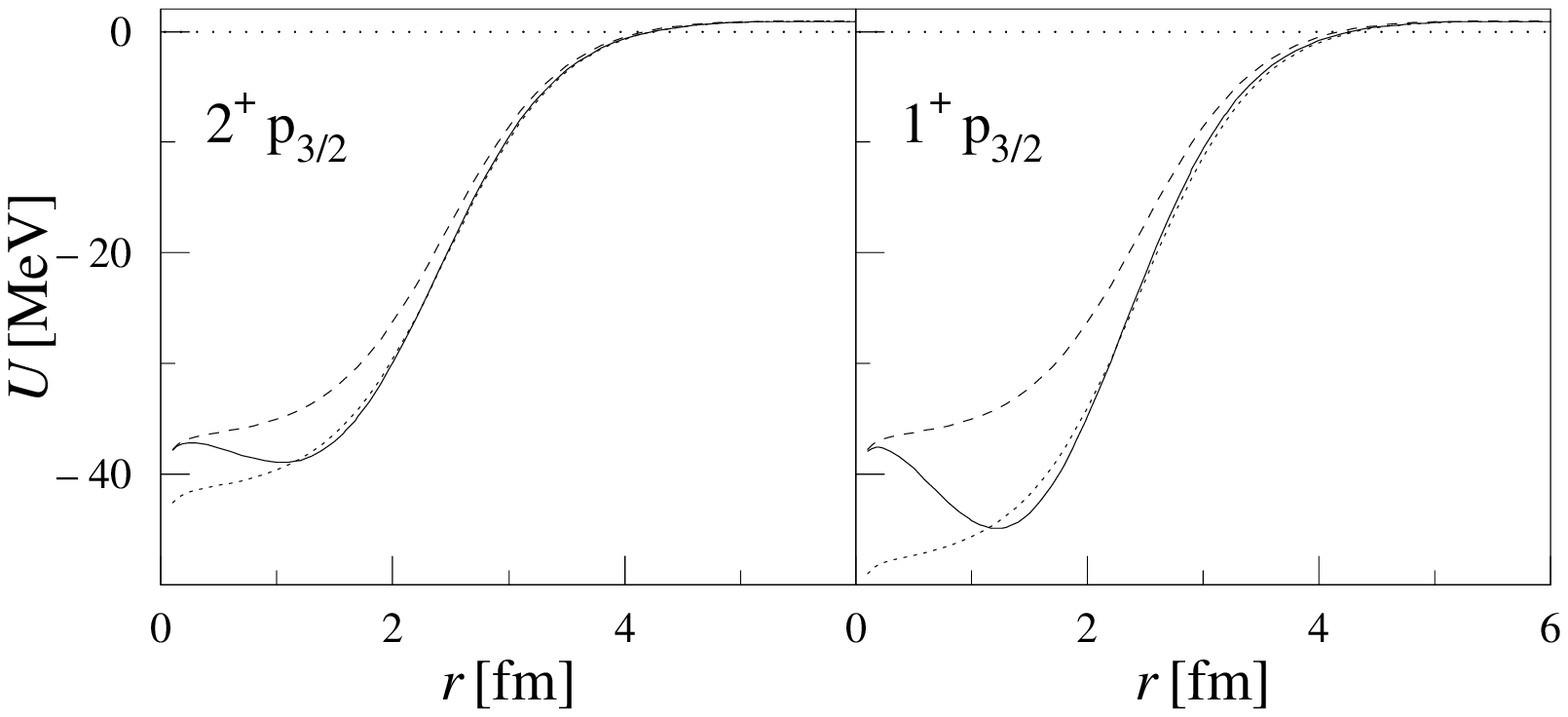,height=8cm}}
\vskip 1truecm
\caption{Finite-depth average s.p.\ potentials used to generate the radial s.p.\ wave
functions for bound states and resonances. \\
(i) Plot on the l.h.s.\ shows the initial potential $U(r)$ (\protect\ref{pot})
(the dashed line), the self-consistent potential $U^{(sc)}(r)$
(the solid line), {\it i.e.},
$U(r)$ which is corrected by the coupling to the continuum of scattering
states, and the equivalent
average potential $U^{(eq)}(r)$ (the dotted line) of the Saxon-Woods
type which yields the proton $1p_{3/2}$ orbit at the same
energy as in the self-consistent potential. The residual
interaction (\protect\ref{force}) parameters are:
$V_{12}^{(0)} = 650\,$MeV$\cdot$fm$^3$, $\alpha = 0.73$.  $U(r)$ is
chosen in such a way that $U^{(sc)}(r)$
yields the $1p_{3/2}$ s.p.\ state at the energy $-137\,$keV,
corresponding to the binding energy of the $2_{1}^{+}$ g.s.\ in
$^{8}\mbox{B}$. The correction to the average potential from the residual
interaction (\protect\ref{force}) is calculated for the $1p_{3/2}$ s.p.\ orbit in
$2^{+}$ states of $^{8}\mbox{B}$. For more details, see
Table~\protect\ref{parameters} and the description in the text.\\
(ii) Plot on the r.h.s.\ shows the same as on l.h.s.\ but for the
$1^{+}$ states of $^{8}\mbox{B}$. The initial potentials on the r.h.s.\ and
l.h.s.\ are identical and $U^{(eq)}(r)$  is reproducing
the position of the $1p_{3/2}$ s.p.\ orbit of $U^{(sc)}(r)$ for
$1^{+}$ states in  $^{8}\mbox{B}$.}
\label{fig1}
\end{figure}
\newpage
\begin{figure}[t]
\centerline{\epsfig{figure=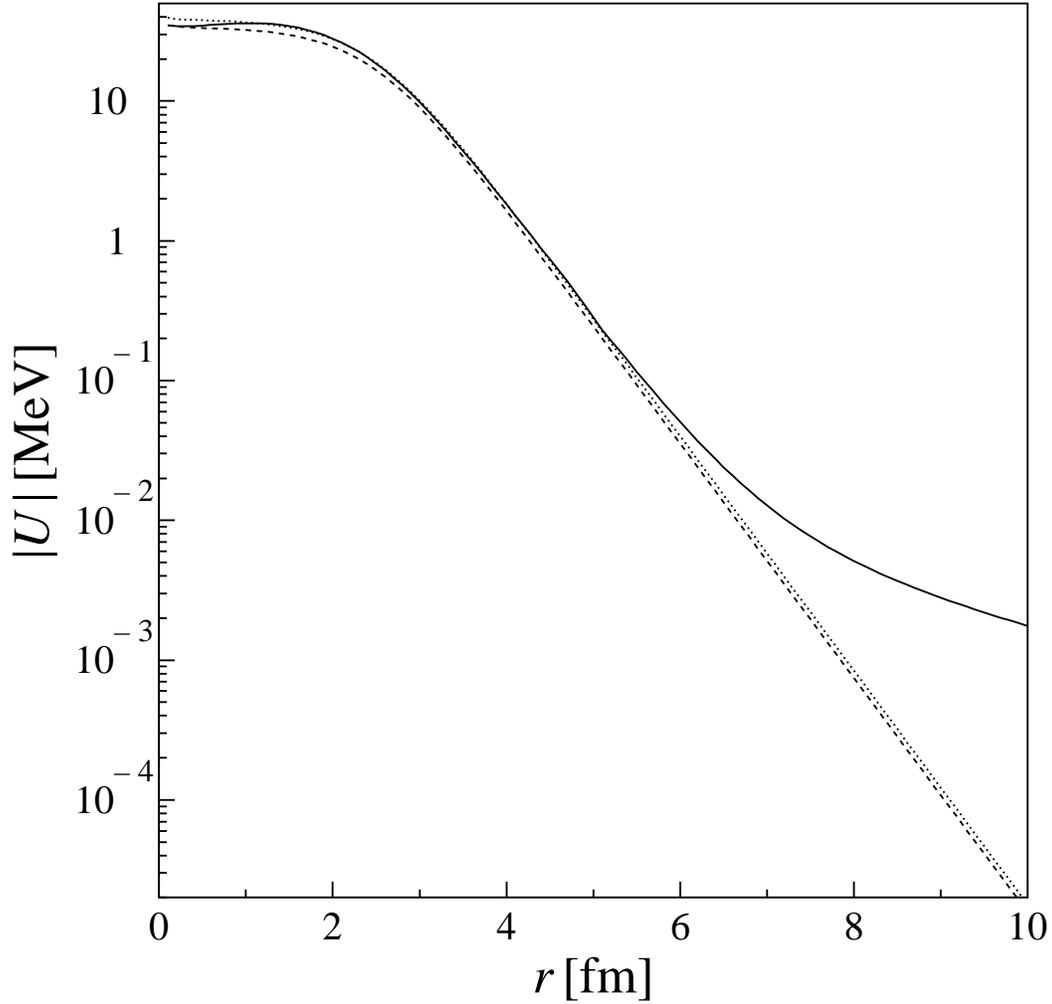,height=14cm}}
\vskip 1truecm
\caption{The initial potential $U(r)$ of Saxon-Woods type (see
(\protect\ref{pot}))
(the dashed line), the self-consistent potential $U^{(sc)}(r)$
(the solid line) and the equivalent Saxon-Woods
average potential $U^{(eq)}(r)$ (the dotted line)
which yields the neutron $1p_{3/2}$ orbit at the same
energy as found in $U^{(sc)}(r)$.
The curves show the absolute value of corresponding potentials in the
logarithmic scale so the characteristic tail for large $r$
can be well seen. $U(r)$ is
chosen in such a way that $U^{(sc)}(r)$ for neutrons obtained from it by
including the diagonal correction term from the residual interaction
(\protect\ref{force}) yields the $1p_{3/2}$ neutron s.p.\ state
in $J^{\pi}=2^{+}$ states of $^{8}\mbox{Li}$ at the energy $-20\,$keV.
For more details, see Table~\protect\ref{parameters} and the description
in the text.}
\label{fig2}
\end{figure}
\newpage
\begin{figure}[t]
\centerline{\epsfig{figure=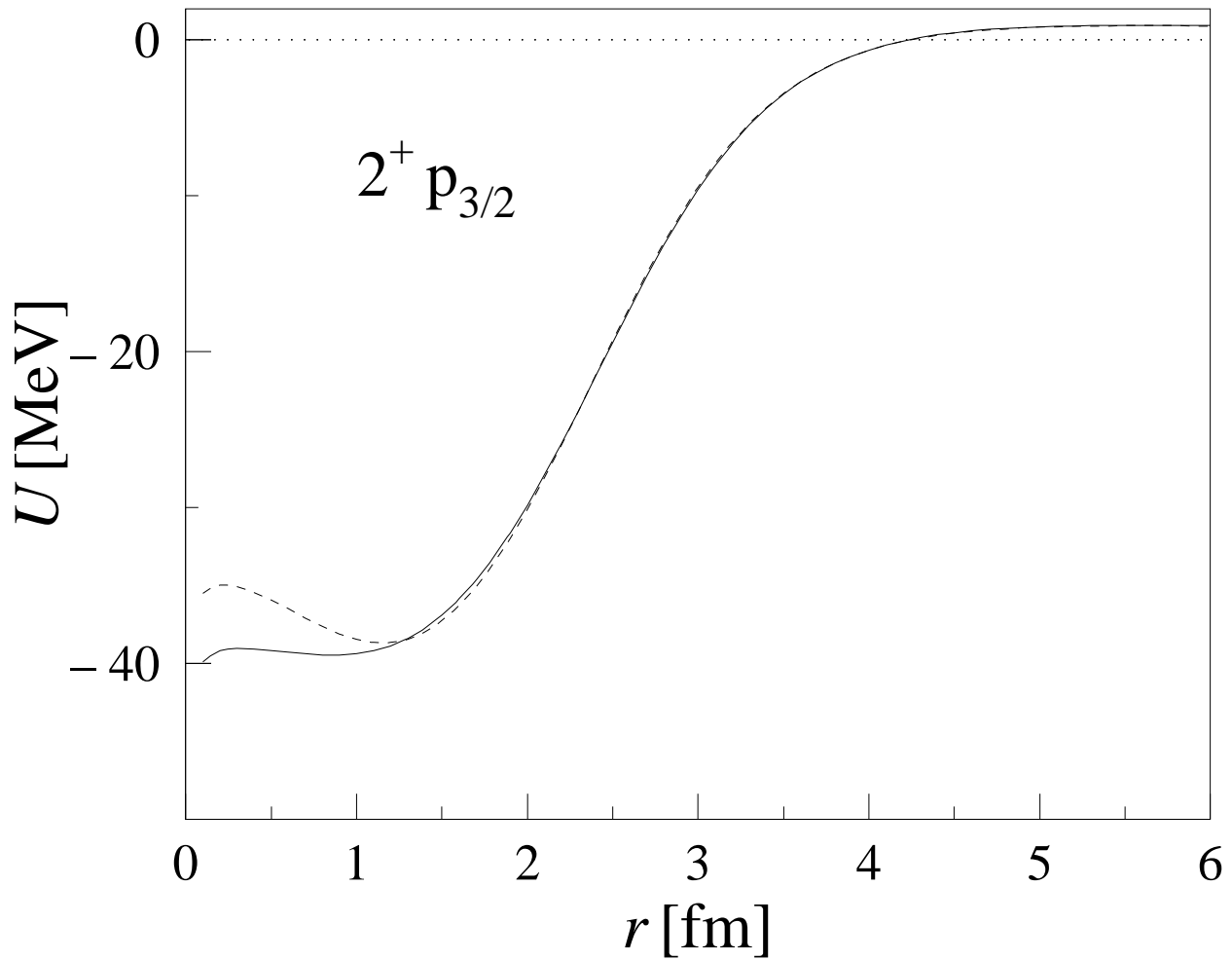,height=14cm}}
\vskip 1truecm
\caption{The self-consistent average potential $U^{(sc)}(r)$ which is
obtained by including the
coupling of states in $Q$ and $P$ subspaces with the residual interaction
(\protect\ref{force}), is plotted for two parameters of the residual force
$\alpha=0.55$ (the solid line)
and $\alpha=0.95$ (the dashed line).
The calculations have been done for the s.p.\ orbit $1p_{3/2}$ and in
$J^{\pi}=2^{+}$ states of  $^{8}\mbox{B}$.
The initial potential in each case has been chosen in such a way
that the corresponding self-consistent potential yields the $1p_{3/2}$
s.p.\ orbit at the same energy of $-137\,$keV. For more details, see
Table~\protect\ref{parameters} and the description in the text.}
\label{fig3}
\end{figure}
\newpage
\begin{figure}[t]
\centerline{\epsfig{figure=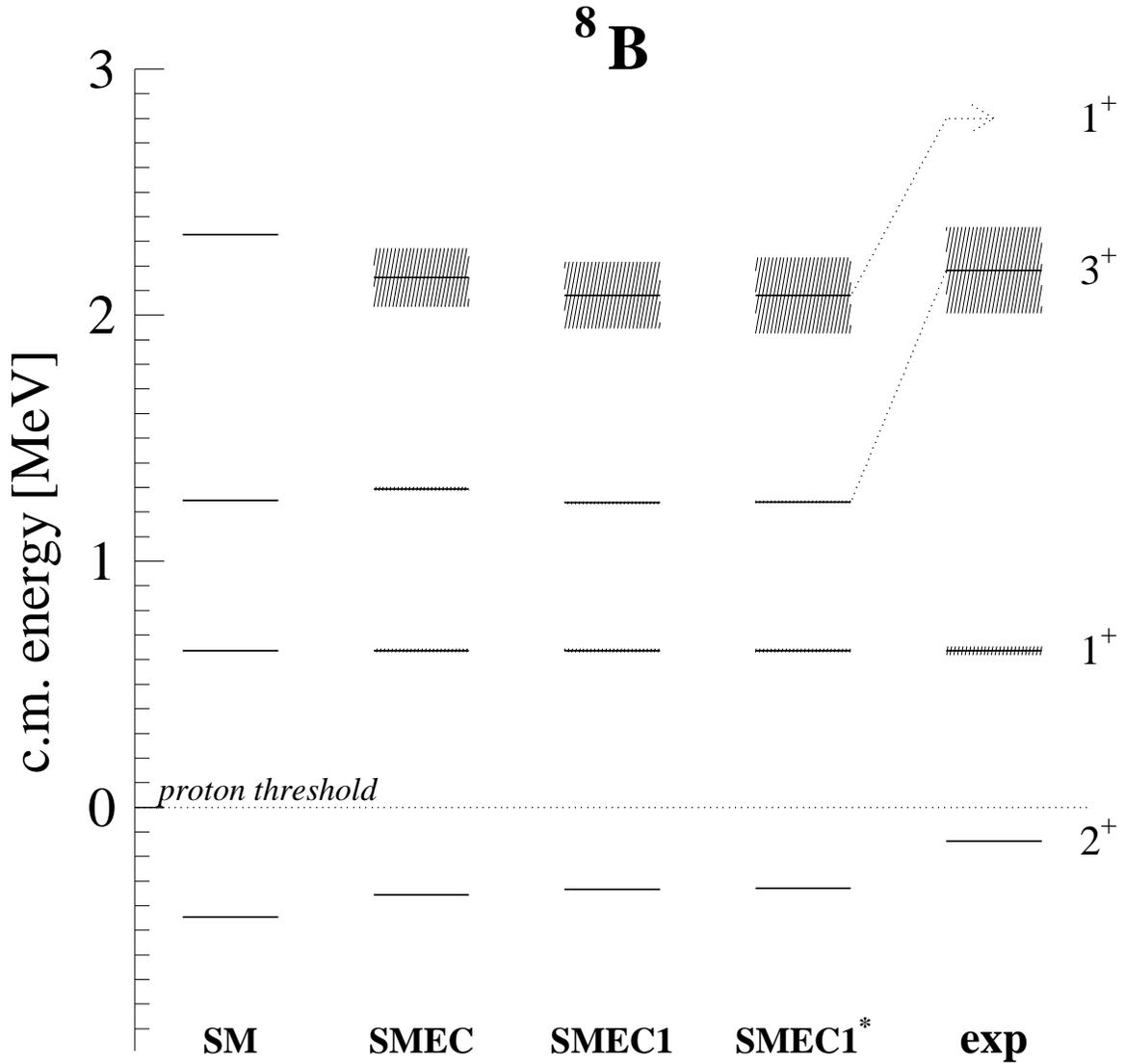,height=17cm}}
\vskip 1truecm
\caption{SM with CK interaction and SMEC in different approximations labelled 'SMEC',
SMEC1' and 'SMEC1$^*$'
	vs. experimental $T=1$ states of    $^{8}\mbox{B}$
 nucleus.  The proton threshold
energy is adjusted to reproduce position of the $1_{1}^{+}$
first excited
state. The shaded regions represent the width of resonance states. For the
details of the calculation, see the description in the text and in the caption
of Table~\protect\ref{parameters}.}
\label{fig4}
\end{figure}
\newpage
\begin{figure}[t]
\centerline{\epsfig{figure=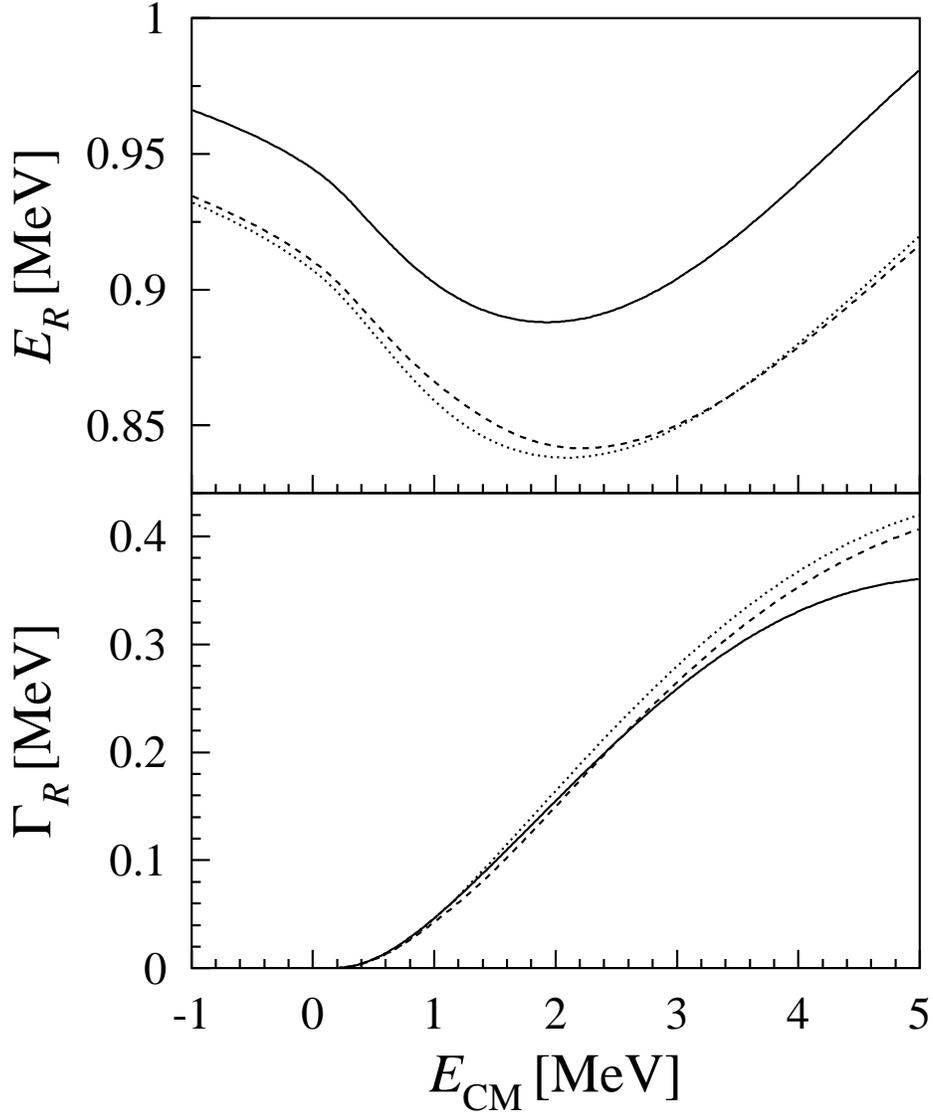,height=16cm}}
\vskip 1truecm
\caption{Energy dependence of the eigenvalue (both real $E_R$ and imaginary
${\Gamma}_R$ parts) of the effective Hamiltonian (\protect\ref{eq2a}) for the
$1_{1}^{+}$ state in $^{8}\mbox{B}$.
The solid line corresponds to the coupling to the g.s.\ of
$^{7}\mbox{Be}$ only. The dashed line corresponds to the inclusion of
coupling to the first excited state in $^{7}\mbox{Be}$ at the energy
predicted by SM with CK interaction \protect\cite{cohen}. The dotted line corresponds
to the inclusion of
coupling to the first excited state in $^{7}\mbox{Be}$ which is placed
at the experimental energy. For more details, see the description
in the text.}
\label{fig6}
\end{figure}
\newpage
\begin{figure}[t]
\centerline{\epsfig{figure=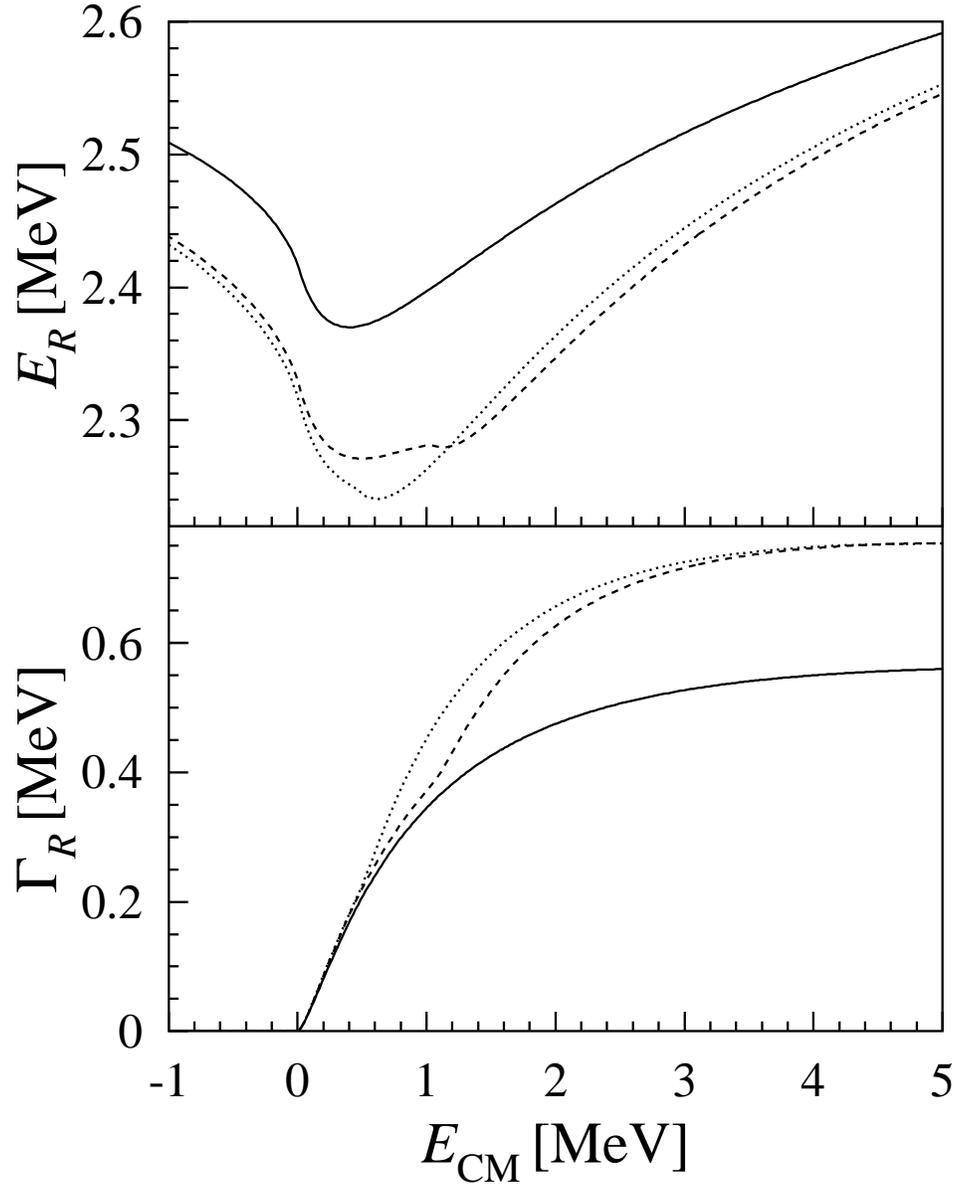,height=17cm}}
\vskip 1truecm
\caption{The same as in Fig. 5 but for the $1_{2}^{+}$ state in $^{8}\mbox{Li}$.
For more details, see the caption of Fig. 5 and the description
in the text.}
\label{fig8}
\end{figure}
\newpage
\begin{figure}[t]
\centerline{\epsfig{figure=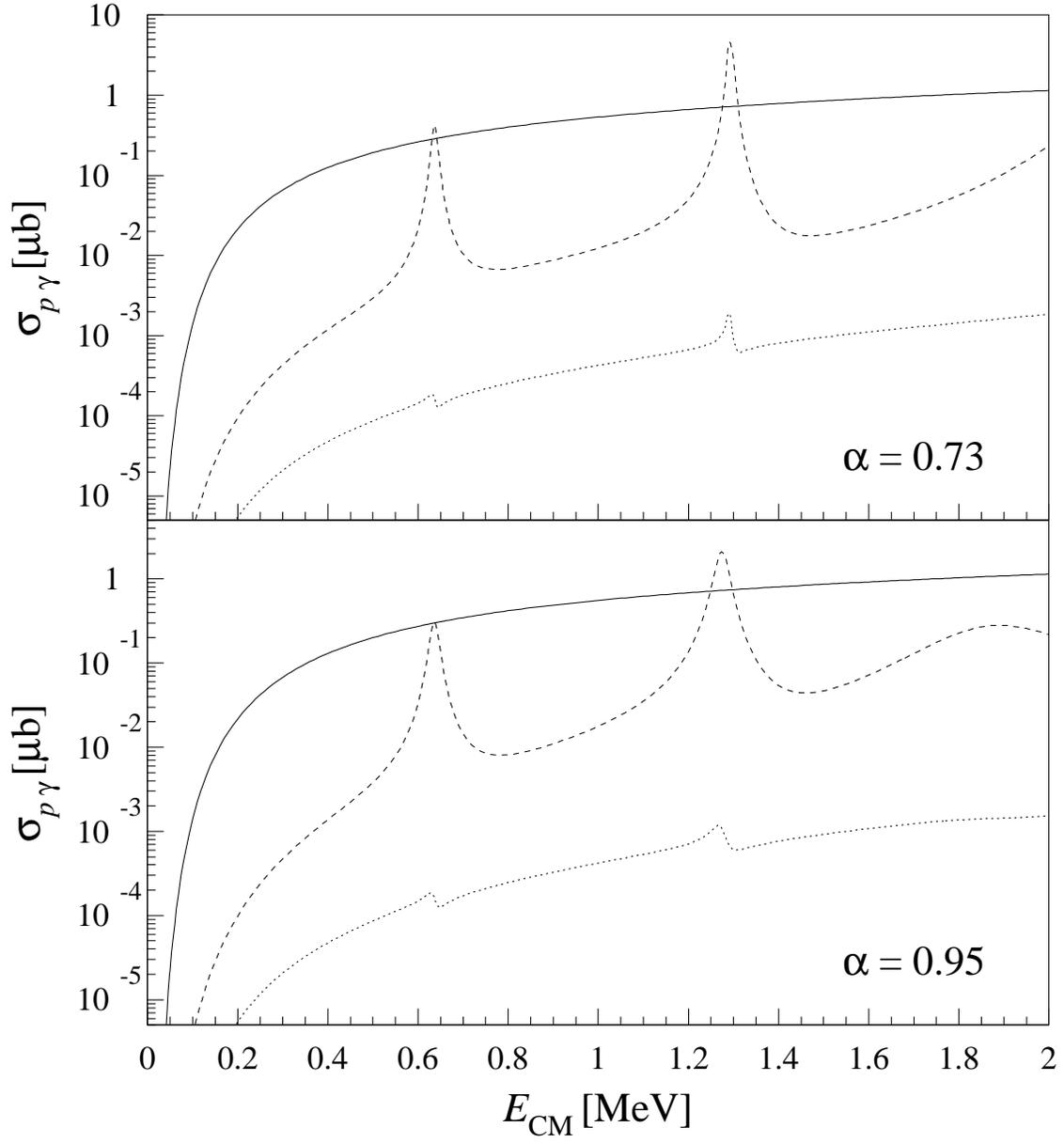,height=17cm}}
\vskip 1truecm
\caption{Multipole contributions to the total capture cross section
of $^{7}\mbox{Be}(p,\gamma)^{8}\mbox{B}$
as a function of the center of mass energy. The
SMEC calculations have been done for
different values of the spin-exchange parameter
$(1-\alpha) = 0.27$ (the upper part of the figure) and
$(1-\alpha) = 0.05$. For other details, see Table~\protect\ref{parameters} and the
discussion in the text.}
\label{fig9}
\end{figure}
\newpage
\begin{figure}[t]
\centerline{\epsfig{figure=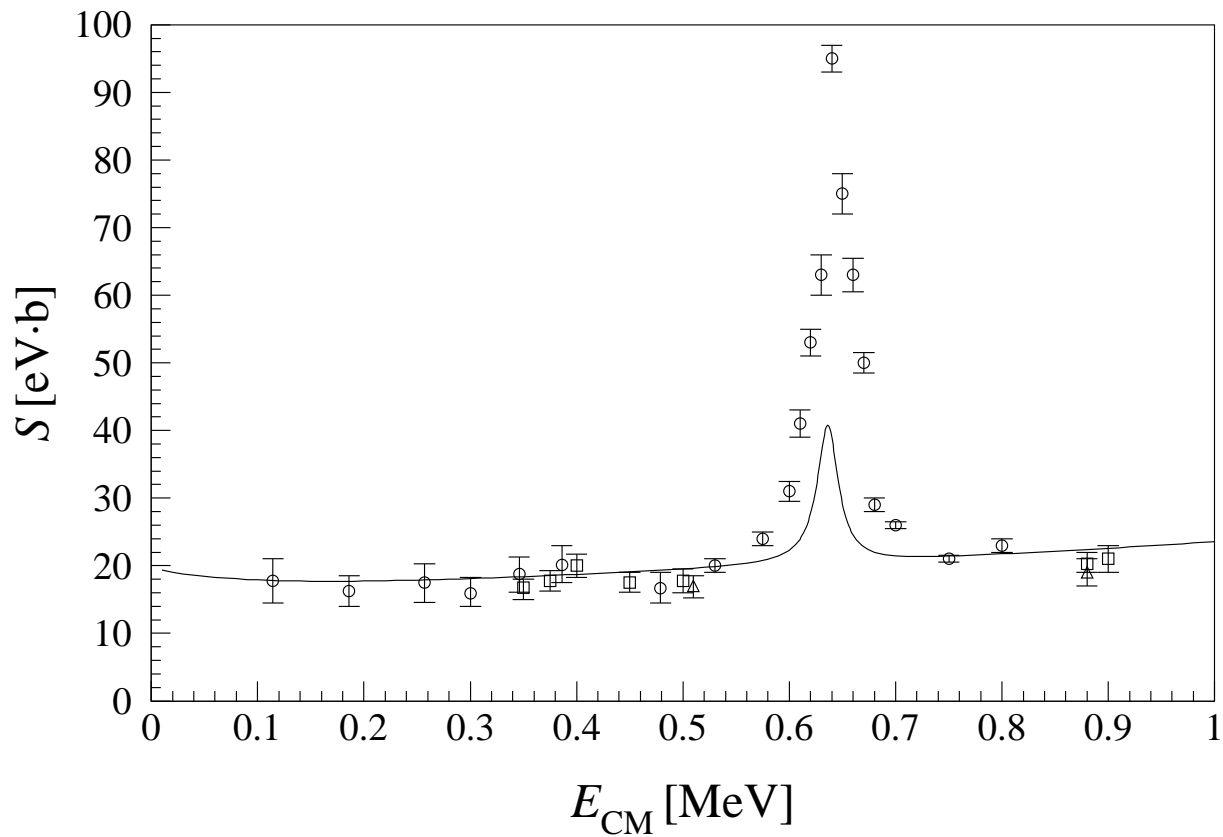,height=12cm}}
\vskip 1truecm
\caption{The astrophysical $S$-factor for the reaction
$^{7}\mbox{Be}(p,\gamma)^{8}\mbox{B}$ is plotted
as a function of the center of mass energy. The
SMEC calculations have been done for
the residual interaction (\protect\ref{force}) the spin-exchange parameter
$(1-\alpha)=0.05$. The experimental points have been taken from Filippone
et al. \protect\cite{filippone} and Hammache et al. \protect\cite{hammache}.}
\label{fig10}
\end{figure}
\newpage
\begin{figure}[t]
\centerline{\epsfig{figure=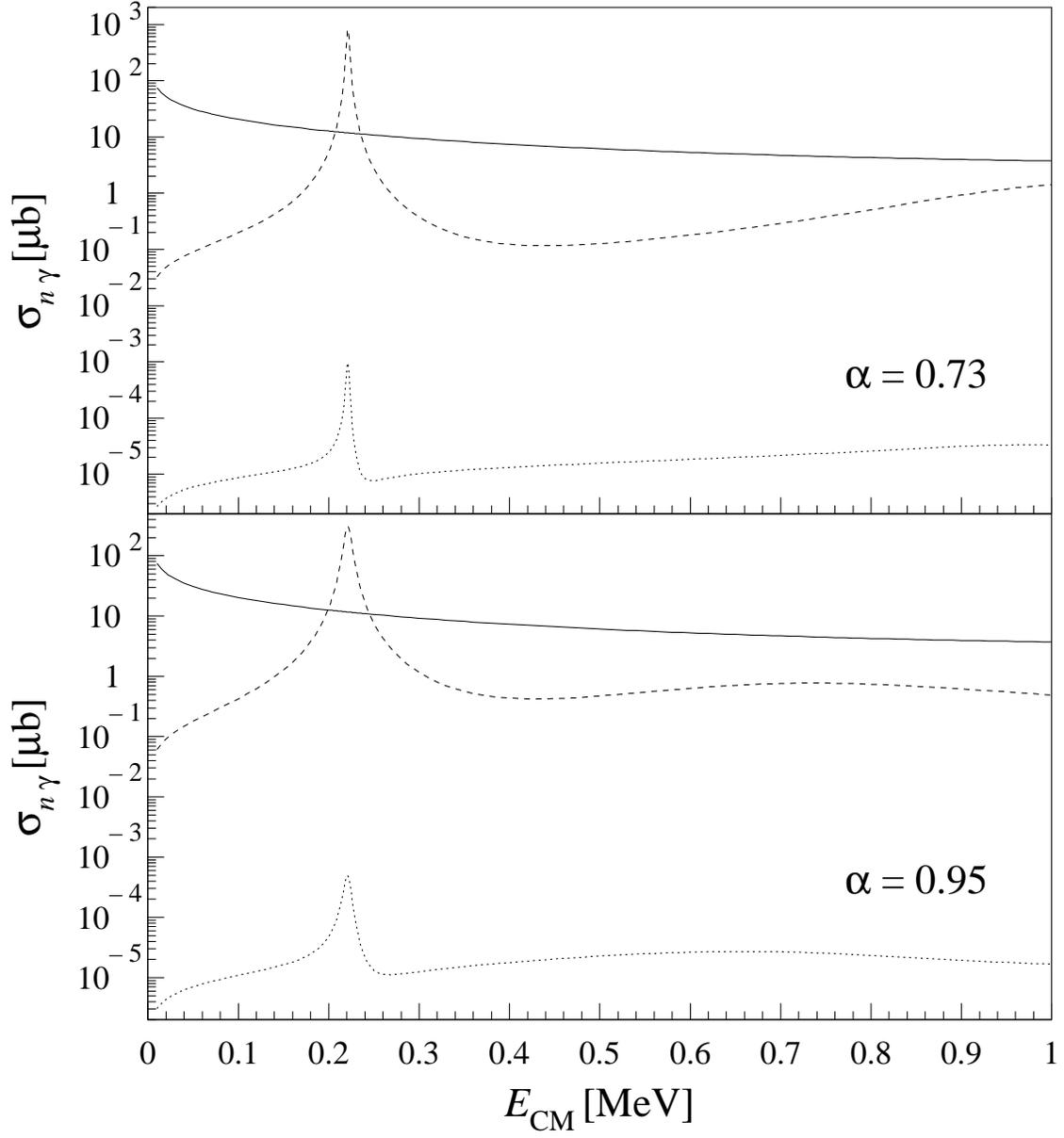,height=17cm}}
\vskip 1truecm
\caption{Multipole contributions to the capture cross section to the
g.s.\ ($J^{\pi}=2_{1}^{+}$) of $^{8}\mbox{Li}$ in the reaction
$^{7}\mbox{Li}(n,\gamma)^{8}\mbox{Li}$
are plotted as a function of c.m.\ energy. The
SMEC calculations have been done with the spin-exchange parameter
$(1-\alpha) = 0.27$ (the upper part of the figure)
and $(1-\alpha)=0.05$. For other details see the discussion in the text.}
\label{fig11}
\end{figure}
\newpage
\begin{figure}[t]
\centerline{\epsfig{figure=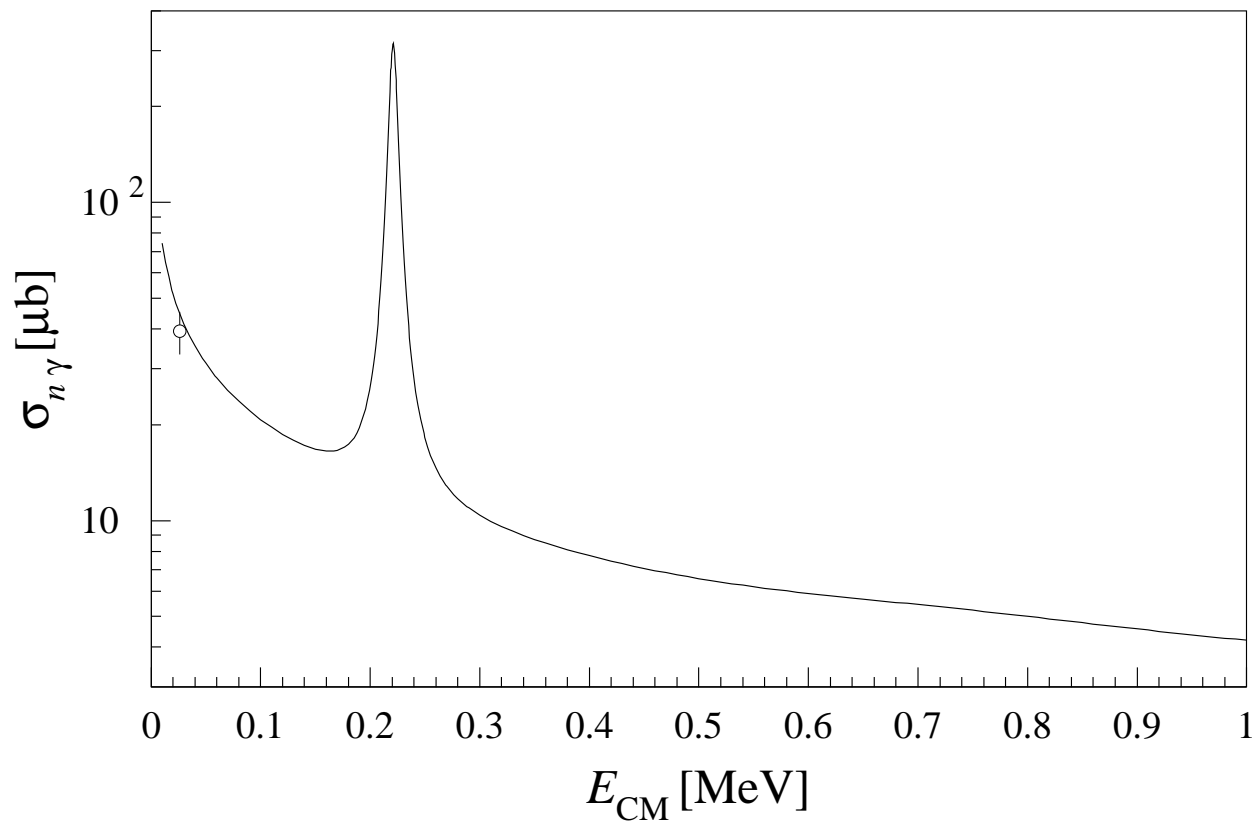,height=12cm}}
\vskip 1truecm
\caption{The cross-section for the reaction
$^{7}\mbox{Li}(n,\gamma)^{8}\mbox{Li}$ is plotted
as a function of c.m.\ energy. The
SMEC calculations have been done for the spin-exchange parameter
$(1-\alpha) = 0.05$. The experimental point is taken from Nagai et al.
\protect\cite{nagai}.}
\label{fig12}
\end{figure}

\end{document}